\documentclass[12pt]{article}

\usepackage{amsfonts}
\usepackage{amsmath}
\usepackage{amsthm}
\usepackage{cite}
\usepackage{epsfig}
\usepackage{latexsym}
\usepackage{paralist}
\usepackage{fancyhdr}
\usepackage{graphicx}
\numberwithin{equation}{section}
\usepackage[vcentermath]{youngtab}
\usepackage{young}
\usepackage{ytableau}
\usepackage{etex}
\usepackage{braket}
\usepackage{float}
\usepackage{extarrows}
\usepackage{autobreak}

\usepackage{pict2e}   
\usepackage{animate}  

\usepackage{multirow}
\usepackage{bigdelim}
\usepackage{fancybox}
\usepackage{cals}
\usepackage{wick}

\def\be{\begin{equation}}
\def\ee{\end{equation}}
\def\bea{\begin{eqnarray}}
\def\eea{\end{eqnarray}}

\renewcommand{\thefootnote}{\fnsymbol{footnote}}

\begin{document}

\hfuzz=100pt
\title{{\Large \bf{Generalized Giveon-Kutasov duality}  }}
\date{}
\author{ Keita Nii$^a$\footnote{keita.nii@yukawa.kyoto-u.ac.jp}
}
\date{\today}

\maketitle

\thispagestyle{fancy}
\cfoot{}
\renewcommand{\headrulewidth}{0.0pt}

\vspace*{-1cm}
\begin{center}
$^{a}${{\it Center for Gravitational Physics}}
\\{{\it Yukawa Institute for Theoretical Physics, Kyoto University
}}
\\ {{\it Kitashirakawa Oiwake-cho, Sakyo-Ku, Kyoto, Japan}}  

\end{center}

\begin{abstract}
We generalize the Giveon-Kutasov duality by adding possible Chern-Simons interactions for the $U(N)$ gauge group. Some of the generalized dualities are known in the literature and many others are new to the best of our knowledge. The dualities are connected to the non-supersymmetric bosonization duality via mass deformations. For $N=1$, there are an infinite number of magnetic-dual theories.   
\end{abstract}

\renewcommand{\thefootnote}{\arabic{footnote}}
\setcounter{footnote}{0}

\newpage
\tableofcontents 

\newpage

\section{Introduction}
Duality is a powerful tool to study the low-energy dynamics of strongly-coupled gauge theories. In supersymmetric theories with four supercharges, this is known as Seiberg(-like) dualities \cite{Seiberg:1994bz, Seiberg:1994pq}. By using the dualities, we can study the non-perturbative aspects of the electric gauge theories through the semi-classical analysis of the magnetic-dual descriptions. Recently, the 3d Seiberg-like dualities have been extremely developed and generalized. One of the reasons for this development is that the quantum structure of the Coulomb moduli space is deeply understood and we have been able to derive various 3d dualities from 4d dualities \cite{Aharony:2013dha, Aharony:2013kma}.

The Seiberg duality was first proposed in 4d and generalized by altering the gauge group and introducing various matter fields \cite{Intriligator:1995ax, Kutasov:1995ve, Kutasov:1995np, Intriligator:1995id, Intriligator:1995ne}. After the development in 4d, the similar dualities were also proposed in diverse dimensions. In 3d, there is a new type of the Seiberg duality, which has no 4d analog. One of these dualities is known as the Giveon-Kutasov duality \cite{Giveon:2008zn} which is a supersymmetric Chern-Simons (CS) duality. This duality was also generalized by including various matter fields with various gauge groups \cite{Niarchos:2008jb, Niarchos:2009aa, Kapustin:2011vz, Kapustin:2011gh, Benini:2011mf, Willett:2011gp, Aharony:2014uya, Hwang:2012jh, Hwang:2015wna, Park:2013wta, Hwang:2011qt, Benini:2017dud, Dimofte:2017tpi, Bashmakov:2018ghn, Benvenuti:2018bav, Amariti:2018wht}.

In this paper, we will further generalize the $U(N)_k$ Giveon-Kutasov (GK) duality. The conventional GK duality deals with the $U(N)_{k}$ gauge group where the CS levels for the abelian and non-abelian subgroups are identical. In this paper, we consider more generic CS levels for the $U(N)$ gauge group. Namely, we introduce two different CS terms for the abelian and non-abelian subgroups. This generalization has been considered in non-supersymmetric CS theories \cite{Radicevic:2016wqn, Gur-Ari:2015pca, Aharony:2015mjs, Hsin:2016blu, Benini:2017aed} and some of the supersymmetric theories \cite{Choi:2018ohn}. This paper will propose the 3d $\mathcal{N}=2$ duality for the $U(N)_{k,k+nN}$ gauge group.

The rest of this paper is organized as follows: 
In Section 2, we will propose the generalized Giveon-Kutasov duality and discuss the connection with the non-supersymmetric bosonization duality by breaking supersymmetry with mass terms for the gaugino field. 
Section 3, 4, and 5 are devoted to the detailed analysis of the proposed duality by taking special values of $n$, which reduces to the known dualities. 
In Section 6, we will give some concrete examples by focusing on the $U(2)$ and $U(1)$ gauge groups with small flavors. As a consistency check, we will compute the superconformal indices for some cases.
In Section 7, we will summarize our findings and discuss possible future directions.

\section{Generalized Giveon-Kutasov duality}
In this section, we will propose the generalized Giveon-Kutasov duality. The electric theory is a 3d $\mathcal{N}=2$ $U(N)_{k,k+nN}$ gauge theory with $F$ fundamental flavors, which is a vector-like theory. The bare Chern-Simons (CS) level for the $SU(N)$ subgroup is $k \neq 0$ whereas the CS level for the abelian factor is $k+nN$, where $k$ and $n$ are integers. For these CS levels, we will use the notation adopted, for example, in \cite{Aharony:2015mjs, Hsin:2016blu, Radicevic:2016wqn}. Notice that we can introduce two independent CS factors for the $U(N)$ gauge group since $U(N) =SU(N) \times U(1)/\mathbb{Z}_N$. The $n=0$ case corresponds to the conventional Giveon-Kutasov duality proposed in \cite{Giveon:2008zn}. The theory has the $SU(F)\times SU(F) \times U(1)_A \times U(1)_T \times U(1)_R$ symmetry, where $U(1)_T$ denotes a topological $U(1)$ symmetry associated with the overall $U(1) \subset U(N)$ gauge group. The quantum numbers of the elementary fields are summarized in Table \ref{GGKele}. Since the gauge group is unitary (not special unitary), there is no baryonic branch in the moduli space of vacua. The Higgs branch is then parametrized by the meson operator $M:=Q \tilde{Q}$. The detailed analysis of the Coulomb moduli space will be given in the following discussion.  

\begin{table}[H]\caption{The 3d $\mathcal{N}=2$ $U(N)_{k,k+nN}$ gauge theory with $F(\,{\tiny \protect\yng(1)}_{\, 1}+ {\tiny \overline{\protect\yng(1)}}_{\,-1})$} 
\begin{center}
\scalebox{1}{
  \begin{tabular}{|c||c|c|c|c|c|c| } \hline
  &$U(N)_{k,k+nN}$&$SU(F)$&$SU(F)$&$U(1)_A$&$U(1)_T$&$U(1)_R$ \\ \hline
$Q$& ${\tiny \yng(1)}_{\, 1}$&$\tiny \yng(1)$&1&$1$&0&$0$ \\  
$\tilde{Q}$& ${\tiny \overline{\yng(1)}}_{\, -1}$&1&$\tiny \yng(1)$&1&0&0  \\ \hline
$M:=Q\tilde{Q}$&1&${\tiny \yng(1)}$&$\tiny \yng(1)$&$2$&0&$0$  \\  \hline
  \end{tabular}}
  \end{center}\label{GGKele}
\end{table}

The dual theory is easily obtained by slightly modifying the non-supersymmetric duality proposed in \cite{Radicevic:2016wqn}.
The magnetic description is given by a 3d $\mathcal{N}=2$ $U(1)_{n+1} \times U(\tilde{N})_{-k,-k+\tilde{N}}$ gauge theory with $F$ fundamental (dual) flavors and a meson singlet $M$, where $\tilde{N}=F+k-N$. The first $U(1)$ gauge group is a topological $U(1)$ symmetry associated with the $U(\tilde{N})_{-k,-k+\tilde{N}}$ gauge group. As a result, there is a level-$1$ mixed CS term between $U(1)_{n+1}$ and $U(\tilde{N})_{-k,-k+\tilde{N}}$. Generally, we can choose the level of the mixed CS term to be any integer. However, we will see that the level-$1$ mixed CS term correctly produces the desired duality. As usual, the magnetic theory has a tree-level superpotential
\begin{align}
W_{mag} = Mq\tilde{q}.
\end{align}
Table \ref{GGKmag} summarizes the quantum numbers of the magnetic fields. Note that the magnetic matter fields are not charged under the topological (gauged) $U(1)_{n+1}$ symmetry. The charge assignment is fixed as in Table \ref{GGKmag} because the theory is vector-like and there is a swapping symmetry between $Q$ and $\tilde{Q}$ (or $q$ and $\tilde{q}$ on the magnetic side). In the following sections, we will see the validity of our duality proposal for various values of $n$.

\begin{table}[H]\caption{The magnetic $\stackrel{+1~~~~~~~~~~~~~~~~~~~~~~~~~~}{\protect\wick{1}{<1 U(1)_{n+1} \times >1 U(F+k-N)_{-k,F-N} }}$ gauge theory dual to Table \ref{GGKele}} 
\begin{center}
\scalebox{1}{
  \begin{tabular}{|c||c|c|c|c|c|c| } \hline
  &$\stackrel{+1~~~~~~~~~~~~~~~~~~~~~~~~~~}{\protect\wick{1}{<1 U(1)_{n+1} \times >1 U(F+k-N)_{-k,F-N} }}$&$SU(F)$&$SU(F)$&$U(1)_A$&$U(1)_T$&$U(1)_R$ \\ \hline
$q$& ${\tiny \yng(1)}_{\, 1}$&$\tiny \overline{\yng(1)}$&1&$-1$&0&$1$ \\  
$\tilde{q}$& ${\tiny \overline{\yng(1)}}_{\, -1}$&1&$\tiny \overline{\yng(1)}$&$-1$&0&1  \\ 
$M$&1&${\tiny \yng(1)}$&$\tiny \yng(1)$&$2$&0&$0$ \\   \hline
  \end{tabular}}
  \end{center}\label{GGKmag}
\end{table}

As a simple test of this duality, we can derive the non-supersymmetric duality proposed in \cite{Radicevic:2016wqn} via supersymmetry-breaking deformations. We start from the supersymmetric duality with $F=0$. On the electric side, we introduce a mass term for the gaugino field, which explicitly breaks supersymmetry. After integrating out the massive gaugino, the resulting theory is a (non-supersymmetric) 3d $U(N)_{k',k'+(n+1)N}$ pure CS theory, where $k' = k-N$ is a shifted CS level. On the magnetic side, the gauge group is $U(1)_{n+1} \times U(k-N)_{-k, -k+(k-N)}$ for $F=0$. After integrating out the magnetic gaugino, we obtain a $U(1)_{n+1} \times U(k')_{-N,-N}$ pure CS gauge theory. Notice that the $U(1)$ CS level is not shifted by the massive gaugino field. This is precisely the duality proposed in \cite{Radicevic:2016wqn}. For $F \neq 0$, on the electric side, we introduce masses for the gaugino and the matter fermions in the chiral multiplets, which leads to a $U(N)_{\tilde{k},\tilde{k}+(n+1)N}$ gauge theory with 2F scalar fields, where $\tilde{k}=k-N+F$. On the magnetic side, we assume that the mass terms for the electric matter fermions are mapped to the mass terms for the matter sfermions. The resulting theory is a $U(1)_{n+1} \times U(\tilde{k})_{F-N,F-N}$ gauge theory with $2F$ fermions. This is again the non-supersymmetric duality proposed in \cite{Radicevic:2016wqn}. We can regard this flow of the dualities as a quick consistency check although this is not a rigorous derivation.

For $n=0$, we can recover the conventional Giveon-Kutasov duality for the $U(N)_{k,k}$ gauge group, where the Chern-Simons levels for the abelian and non-abelian parts are equal. In the literature, this case is simply written as a $U(N)_k$ gauge group. On the magnetic side, the gauge group becomes $U(1)_{1} \times U(F+k-N)_{-k, F-N}$ with a level-1 mixed CS term. Since the absolute value of the $U(1)_{top}$ CS level is unity, we can integrate over the associated vector multiplet. By taking a linear combination of the two $U(1)$ gauge fields, the theory becomes a 3d $\mathcal{N}=2$ $U(1)_1 \times U(F+k-N)_{-k,-k}$ gauge theory without a mixed CS term between these two new $U(1)$ subgroups. The matter sector is only charged under the $U(F+k-N)_{-k,-k}$ gauge group. Therefore, we can simply turn off the topological $U(1)_1$ gauge dynamics. In this way, we reproduce the conventional Giveon-Kutasov duality \cite{Giveon:2008zn}.

Next, we consider applying the duality transformation twice to the original theory, which must go back to the original theory itself. By applying the duality to the electric theory just once, the first dual is given by the $\stackrel{+1~~~~~~~~~~~~~~~~~~~~~~~~~~}{\wick{1}{<1 U(1)_{n+1} \times >1 U(F+k-N)_{-k,F-N} }}$ gauge theory as in Table \ref{GGKmag}. By further applying the duality transformation to the $U(F+k-N)$ gauge group, the dual is given by the $U(1)_{n+1} \times U(1)_0 \times U(N)_{k,k-N}$ gauge theory with the superpotential
\begin{align}
W_{second\,mag} =MN +Nq'\tilde{q}',
\end{align}
where $N$ is identified with the dual meson $q \tilde{q}$. By using the $F$-term conditions for $M$ and $N$, we find that these two mesons are massive and the superpotential vanishes. The level-1 mixed CS term is introduced between the $U(1)_{n+1}$ and $U(1)_0$ subgroups while the mixed CS term between the $U(1)_0$ and $U(1) \subset U(N)$ subgroups is $-1$. Since the second $U(1)_0$ gauge group has only the BF interaction, we can integrate over the corresponding vector multiplet. The resulting gauge group becomes $U(N)_{k,k+nN}$ and we reproduce the electric gauge theory. This also serves as a consistency check of our duality.

Finally, we will see the matching of the quantum Coulomb moduli space under the duality. In order to make the discussion simple, we will flip the sign of $n$ (inserting $n \rightarrow -n$) and take $k= nC$ (we here assume $C \le N$). The duality relates the $U(N)_{nC,n(C-N)}$ gauge theory to the $U(1)_{-n+1} \times U(F+nC-N)_{-nC,-nC+(F+nC-N)}$ magnetic dual. On the electric side, we can consider the following Coulomb branch 
\begin{align}
U(N)_{nC,n(C-N)} & \rightarrow \stackrel{-n~~~~~~~~~~~~~~~~~~~~~~~~~~~~~~}{\protect\wick{1}{<1 U(C)_{nC,0} \times >1 U(N-C)_{nC,nC-n(N-C)} }} \\
{\tiny \yng(1)}_{\,1}  & \rightarrow  ({\tiny \yng(1)}_{\,1}, \mathbf{1}_0)+ (\mathbf{1}_0, {\tiny \yng(1)}_{\,1})\\ 
{\tiny \overline{\yng(1)}}_{\,-1}  & \rightarrow  ({\tiny \overline{\yng(1)}}_{\,-1}, \mathbf{1}_0) +(\mathbf{1}_0, {\tiny \overline{\yng(1)}}_{\,-1}),
\end{align}
where the Coulomb branch is associated with the abelian generator of the $U(1)_C \subset U(C)$ subgroup. This breaking is realized by the vev of the adjoint scalar in the $U(N)$ vector multiplet
\begin{align}
\braket{\phi_{adj.}} =\mathrm{diag.} (\overbrace{v,\cdots,v}^{C},0,\cdots,0).
\end{align}
Depending on the sign of $v$, we introduce two coordinates $X_{\pm}$. Since the $U(1)_C \subset U(C)$ CS term is zero, this flat direction (and the associated $U(1)$ vector multiplet) becomes exactly massless. Due to the mixed CS term between the $U(C)$ and $U(N-C)$ subgroups, the bare monopoles $X_{\pm}$ are charged under the $U(1) \subset U(N-C)$ subgroup \cite{Affleck:1982as, Aharony:1997bx, Intriligator:2013lca}. In order to describe the moduli space in a gauge-invariant way, we need to define the dressed monopole operators \cite{Csaki:2014cwa, Amariti:2015kha, Nii:2018bgf}
\begin{align}
X_{+ d} &:= X_+  \left[ (\mathbf{1}_0, {\tiny \overline{\yng(1)}}_{\,-1})^{N-C} \right]^{nC} \sim X_+ \tilde{Q}^{nC(N-C)}  \\
X_{- d} &:= X_-  \left[  (\mathbf{1}_0, {\tiny \yng(1)}_{\,1})^{N-C} \right]^{nC} \sim X_- Q^{nC(N-C)} 
\end{align}

On the magnetic side, this flat direction corresponds to the following Coulomb branch 
\begin{align}
U(1)_{-n+1} & \times U(F+nC-N)_{-nC,-nC+(F+nC-N)} \nonumber \\
& \rightarrow U(1)_{-n+1} \times U((n-1)C)_{-nC,-C} \times U(F+C-N)_{-nC,-nC+(F+C-N)},
\end{align}
where the level-1 mixed CS terms are introduced for all the combinations of the $U(1)$ gauge symmetries. We denote the monopole operators for the topological $U(1)_{-n+1}$ subgroup by $\tilde{\mathcal{U}}_{\pm}$ and the monopoles for the $U(1) \subset U((n-1)C)$ subgroup by $\tilde{X}_{\pm}$. The above breaking is induced by the vevs of $\tilde{X}_{\pm}$. Since all the subgroups obtain the non-zero CS terms, we cannot independently turn on the corresponding Coulomb flat directions. However, we can turn on the following monopole operators
\begin{align}
\tilde{\mathcal{U}}_+^{C} \tilde{X}_+,~~~~~~~~~~~~\tilde{\mathcal{U}}_-^C \tilde{X}_-.
\end{align}
This is equivalent to simultaneously turning on the non-zero vevs for the $U(1)_{n+1}$ and $U((n-1)C)$ adjoint scalars. One can see that these combinations are neutral under the $U(1)_{n+1} \times U((n-1)C)$ subgroup, which implies that there is no CS term for the $U(1)_{n+1} \times U((n-1)C)$ subgroup along the combined Coulomb branch. Due to the mixed CS terms, these are only charged under the $U(1) \subset U(F+C-N)$ gauge symmetry. Therefore, we need to define the dressed monopole operators 
\begin{align}
X_{- d} &\leftrightarrow \tilde{\mathcal{U}}_+^{C} \tilde{X}_+ \left[  (\mathbf{1}_0, {\tiny \yng(1)}_{\,1})^{F+C-N} \right]^{nC} \sim  \tilde{\mathcal{U}}_+^{C} \tilde{X}_+ q^{nC(F+C-N)}\\
X_{+ d} &\leftrightarrow \tilde{\mathcal{U}}_-^{C} \tilde{X}_-  \left[ (\mathbf{1}_0, {\tiny \overline{\yng(1)}}_{\,-1})^{F+C-N} \right]^{nC} \sim  \tilde{\mathcal{U}}_-^{C} \tilde{X}_-  \tilde{q}^{nC(F+C-N)},
\end{align}
which describe the Higgs-Coulomb mixed branch. 
From the quantum numbers of these dressed composites, we obtain the above identification. 
See Table \ref{QNofbaryonmonopole} for the quantum numbers of the bare and dressed Coulomb branch coordinates.

\begin{table}[H]\caption{The quantum numbers of the dressed Coulomb branch operators} 
\begin{center}
\scalebox{1}{
  \begin{tabular}{|c||c|c| } \hline
&$U(1)_A$&$U(1)_R$ \\ \hline
$X_{\pm}$&$-CF$& $C(F-N+C)$ \\ \hline 
$\tilde{\mathcal{U}}_{\pm}^{C} \tilde{X}_{\pm}$&$(n-1)CF$&$-(n-1)C(F+C-N)$  \\   \hline
$X_{\pm d}$&$nC(N-C)-CF$&$C(F-N+C)$ \\  \hline 
  \end{tabular}}
  \end{center}\label{QNofbaryonmonopole}
\end{table}

\section{$n=\infty$: $SU(N)_k$ Giveon-Kutasov duality}
For the $n=\infty$ limit, we can forget about the overall $U(1)$ gauge dynamics since the $U(1)$ CS interaction becomes weaker and weaker in this limit. Thus, we obtain the $SU(N)_k$ Giveon-Kutasov duality with vector-like matter \cite{Aharony:2013dha, Park:2013wta}. To be more specific, the electric description becomes a 3d $\mathcal{N}=2$ $SU(N)_k$ gauge theory with $F$ (anti-)fundamental flavors.  
On the magnetic side, we can similarly drop off the topological $U(1)_{n+1}$ gauge dynamics and obtain the $U(F+k-N)_{-k,F-N}$ gauge theory. On the electric side, the Higgs branch becomes rich since the $SU(N)$ gauge group allows the (anti-)baryon operators $B:=Q^N$ and $\bar{B}:=\tilde{Q}^N$, where the color indices are contracted by an $SU(N)$ epsilon tensor. As studied in \cite{Aharony:2013dha, Park:2013wta, Aharony:2014uya, 1794498}, there is no Coulomb branch in the electric moduli space. This is because we cannot have any Coulomb flat direction where the tree-level CS term is turned off. Table \ref{SUk_ele} summarizes the quantum numbers of the electric moduli coordinates. 

\begin{table}[H]\caption{The 3d $\mathcal{N}=2$ $SU(N)_{k}$ gauge theory with $F({\tiny \protect\yng(1)}+ {\tiny \overline{\protect\yng(1)}})$} 
\begin{center}
\scalebox{1}{
  \begin{tabular}{|c||c|c|c|c|c|c| } \hline
  &$SU(N)_{k}$&$SU(F)$&$SU(F)$&$U(1)_A$&$U(1)_B$&$U(1)_R$ \\ \hline
$Q$& ${\tiny \yng(1)}$&$\tiny \yng(1)$&1&$1$&1&$0$ \\  
$\tilde{Q}$& ${\tiny \overline{\yng(1)}}$&1&$\tiny \yng(1)$&1&$-1$&0  \\ \hline
$M:=Q\tilde{Q}$&1&${\tiny \yng(1)}$&$\tiny \yng(1)$&$2$&0&$0$  \\  
$B:=Q^N$&1&[$N$-th]&1&$N$&$N$&0  \\
$\bar{B}:=\tilde{Q}^N$&1&1&[$N$-th]&$N$&$-N$&0  \\ \hline
  \end{tabular}}
  \end{center}\label{SUk_ele}
\end{table}

The magnetic side is described by a 3d $\mathcal{N}=2$ $U(F+k-N)_{-k,F-N}$ gauge theory with $F$ dual flavors and a meson singlet which is identified with the electric meson $Q\tilde{Q}$ as usual. The dual meson is eliminated from the chiral ring elements by superpotential 
\begin{align}
W_{mag}=Mq \tilde{q}.
\end{align}
In order to see the matching of the baryonic branch under the duality map, we need to consider the following Coulomb branch 
\begin{align}
U(F+k-N)_{-k,F-N} & \rightarrow \, \stackrel{+1~~~~~~~~\,~~~~~~~}{\protect\wick{1}{<1 U(C)_{-k,-k+C} \times >1 U(P)_{-k,-k+P} }} \\
{\tiny \yng(1)}_{\,1}  & \rightarrow  ({\tiny \yng(1)}_{\,1}, \mathbf{1}_0)+ (\mathbf{1}_0, {\tiny \yng(1)}_{\,1})\\ 
{\tiny \overline{\yng(1)}}_{\,-1}  & \rightarrow  ({\tiny \overline{\yng(1)}}_{\,-1}, \mathbf{1}_0) +(\mathbf{1}_0, {\tiny \overline{\yng(1)}}_{\,-1}),
\end{align}
where the Coulomb branch is associated with the $U(1)_C \subset U(C)$ subgroup and its coordinate $\tilde{X}^{bare, \pm}_{U(C) \times U(P)}$ can be constructed by dualizing the $U(1)_C \subset U(C)$ vector multiplet into a chiral superfiled. The CS levels are decomposed as indicated above. Note that there is also a mixed CS term between the two abelian subgroups. Since we are studying the flat direction of the magnetic Coulomb branch, the CS level $k_{eff}^{U(1)_C}$, which behaves as a topological mass for the $U(1)_C$ vector multiplet, must be zero. Thus, we obtain $C=k$ and $P=F-N$. Due to the mixed CS term, the bare monopole $\tilde{X}^{bare, \pm}_{U(C) \times U(P)}$ obtains a non-zero $U(1)_P$ charge \cite{Intriligator:2013lca, Affleck:1982as, Csaki:2014cwa, Amariti:2015kha, Nii:2018bgf}. The gauge-invariant combinations become
\begin{align}
B & \sim \tilde{X}^{bare, +}_{U(k) \times U(F-N)} (\mathbf{1}_0, {\tiny \yng(1)}_{\,1})^{F-N} \sim \tilde{X}^{bare, +}_{U(k) \times U(F-N)}  q^{F-N}    \\
\bar{B} & \sim  \tilde{X}^{bare, -}_{U(k) \times U(F-N)} (\mathbf{1}_0, {\tiny \overline{\yng(1)}}_{\,-1})^{F-N}  \sim    \tilde{X}^{bare, -}_{U(k) \times U(F-N)} \tilde{q}^{F-N}.
\end{align}
From their quantum numbers listed in Table \ref{SUk_mag}, these are identified with the (anti-)baryon operators of the electric theory. This supports the validity of the duality with $n=\infty$. 

\begin{table}[H]\caption{The magnetic $U(F+k-N)_{-k,F-N}$ gauge theory dual to Table \ref{SUk_ele}} 
\begin{center}
\scalebox{0.98}{
  \begin{tabular}{|c||c|c|c|c|c|c| } \hline
  &$U(F+k-N)_{k}$&$SU(F)$&$SU(F)$&$U(1)_A$&$U(1)_B$&$U(1)_R$ \\ \hline
$q$& ${\tiny \yng(1)}_{\,+1}$&$\tiny \overline{\yng(1)}$&1&$-1$&$\frac{N}{F-N}$&$1$ \\  
$\tilde{q}$& ${\tiny \overline{\yng(1)}}_{\, -1}$&1&$\tiny \overline{\yng(1)}$&$-1$&$-\frac{N}{F-N}$&1  \\
$M$&1&${\tiny \yng(1)}$&$\tiny \yng(1)$&$2$&0&$0$  \\  \hline
$\tilde{X}^{bare, \pm}_{U(k) \times U(F-N)} $&$U(1)_P$: $\mp P$&1&1&$F$&$0$&$N-F$  \\
$B:= \tilde{X}^{bare, +}_{U(k) \times U(F-N)}  q^{F-N} $ &1&[$N$-th]&1&$N$&$N$&0 \\
$\bar{B}:= \tilde{X}^{bare, -}_{U(k) \times U(F-N)} \tilde{q}^{F-N}$&1&1&[$N$-th]&$N$&$-N$&0  \\ \hline
  \end{tabular}}
  \end{center}\label{SUk_mag}
\end{table}

\section{$n=-1$: $U(N)_{k,k-N}$ Generalized GK duality}
For $n=-1$, the electric side becomes a 3d $\mathcal{N}=2$ $U(N)_{k,k-N}$ gauge theory with $F$ fundamental flavors. The Higgs branch is the same as the previous example and parametrized by the meson $M:=Q\tilde{Q}$. We here focus on the Coulomb branch which is spanned by the (dressed) monopole operators. As studied in \cite{Aharony:2014uya, Aharony:2015pla, 1794498}, we have two ways of constructing the dressed Coulomb branch. We first define the Coulomb branch coordinates by using only massless degrees of freedom, which is in harmony with the low-energy picture. After doing that, we will also give another interpretation based on the superconformal indices. 

When the Coulomb branch, denoted by $V_{\pm}$, obtains a non-zero vacuum expectation value, the gauge group is spontaneously broken to
\begin{align}
U(N) & \rightarrow U(C) \times U(P) \\
{\tiny \yng(1)}_{\,1}  & \rightarrow  ({\tiny \yng(1)}_{\,1}, \mathbf{1}_0)+ (\mathbf{1}_0, {\tiny \yng(1)}_{\,1})\\ 
{\tiny \overline{\yng(1)}}_{\,-1}  & \rightarrow  ({\tiny \overline{\yng(1)}}_{\,-1}, \mathbf{1}_0) +(\mathbf{1}_0, {\tiny \overline{\yng(1)}}_{\,-1}),
\end{align}
where the Coulomb branch is associated with the $U(1)_C \subset U(C)$ subgroup and its coordinate can be constructed by dualizing the $U(1)_C \subset U(C)$ vector multiplet into a chiral superfield. The operator $V_{\pm}$ denotes the non-abelian monopole associated to the above breaking \cite{Preskill:1984gd, Weinberg:2012pjx, Csaki:2014cwa, Amariti:2015kha, Nii:2018bgf, 1794498}. Under the topological $U(1)$ symmetry, $V_+$ and $V_-$ are positively and negatively charged, respectively. 
Along the gauge symmetry breaking, the bare CS levels are decomposed as
\begin{align}
k_{eff}^{U(1)_C} =k-C,~~~~~k_{eff}^{U(1)_C,U(1)_P} =-1,
\end{align}
where we only listed the abelian CS terms for our purpose. 
Notice that there is no CS level shift from the chiral multiplets since the theory is vector-like and the level shifts from the fundamental and anti-fundamental quarks completely cancel out each other. Since the Coulomb moduli space is by definition a flat direction of the potential for the adjoint scalar in the vector multiplet, we require that the CS level $k_{eff}^{U(1)_C}$, which acts as a topological mass term for the vector multiplet, should be zero. Therefore, the Coulomb branch with $C=k$ (and then $P=N-k$) only survives and becomes a quantum Coulomb moduli space.

Since there is a mixed CS term between the two abelian factors along the above breaking with $C=k$, the bare monopole operator $V_{\pm}$ obtains a non-zero $U(1)_P \subset U(P)$ charge \cite{Intriligator:2013lca, Csaki:2014cwa, Amariti:2015kha, Nii:2018bgf}. Therefore, we have to define the dressed monopole operators 
\begin{align}
V_{d-}:=V_- ((\mathbf{1}_0, {\tiny \yng(1)}_{\,1}))^{N-k}  \sim V_- Q^{N-k}  \\
V_{d+}:=V_+ ((\mathbf{1}_0, {\tiny \overline{\yng(1)}}_{\,-1}))^{N-k} \sim V_+ \tilde{Q}^{N-k},
\end{align}
where the color indices of $Q^{N-k}$ and $\tilde{Q}^{N-k}$ are contracted by an epsilon tensor of the unbroken $SU(P) \subset U(P)$ gauge group. Since the flavor indices of $Q^{N-k}$ and $\tilde{Q}^{N-k}$ are anti-symmetrized as well, these operators can be regarded as (anti-)baryon-monopoles \cite{Aharony:2013dha, Aharony:2013kma}. The quantum numbers of these operators are summarized in Table \ref{GKelenm1}.

Next, we describe another interpretation of the (dressed) Coulomb branch coordinates in a way that is consistent with the state-counting of the superconformal indices (SCI). 
In the expansion of the superconformal indices, the dressed operators $V_{d\pm}$ are differently observed. Since the SCI is the sum over all the states with possible GNO charges \cite{Bhattacharya:2008bja, Kim:2009wb, Imamura:2011su, Kapustin:2011jm}, the magnetic charge of the non-abelian monopole discussed above cannot appear in the expansion of the SCI. We here define the dressed monopoles based on the SCI. When we insert the monopole operators denoted by $V^{U(N-1)}_{\pm}$, the gauge group is spontaneously broken as 
\begin{align}
U(N) & \rightarrow U(1)_{CB} \times U(N-1) \\
{\tiny \yng(1)}_{\,1}  & \rightarrow  (\mathbf{1}_{1}, \mathbf{1}_0)+ (\mathbf{1}_0, {\tiny \yng(1)}_{\,1})\\ 
{\tiny \overline{\yng(1)}}_{\,-1}  & \rightarrow  (\mathbf{1}_{-1}, \mathbf{1}_0) +(\mathbf{1}_0, {\tiny \overline{\yng(1)}}_{\,-1}) \\
\mathbf{adj.}_{0} & \rightarrow (\mathbf{1}_0,\mathbf{1}_0)+(\mathbf{1}_0, \mathbf{adj.}_{0}) +(\mathbf{1}_{1},{\tiny \overline{\yng(1)}}_{\,-1}) +(\mathbf{1}_{-1},{\tiny \yng(1)}_{\,1}),
\end{align}
where the monopole operator is associated with the $U(1)_{CB}$ subgroup. The representation $\mathbf{adj.}_{0}$ denotes the gaugino field. 
Along this breaking, the bare CS terms are decomposed as
\begin{align}
k_{eff}^{U(1)_{CB}} =k-1, ~~~~~~k_{eff}^{U(1)_{CB},U(1) \subset U(N-1)} =-1.
\end{align}
Due to the (mixed) CS terms $k_{eff}^{U(1)_{CB},U(1) \subset U(N-1)} $, the bare operator $V^{U(N-1)}_{\pm}$ obtains a non-zero $U(1)_{CB}$ and $U(1) \subset U(N-1)$ charges. Therefore, the gauge-invariant states are defined as
\begin{align}
 V_{d-} &:= V^{U(N-1)}_{-} (\mathbf{1}_{1},{\tiny \overline{\yng(1)}}_{\,-1})^{k-1}  (\mathbf{1}_0, {\tiny \overline{\yng(1)}}_{\,-1})^{N-k}   \nonumber  \\
 & \sim  V^{U(N-1)}_{-} W_{\alpha}^{k-1} \tilde{Q}^{N-k} \\
  V_{d+} &:= V^{U(N-1)}_{+} (\mathbf{1}_{-1},{\tiny \yng(1)}_{\,1})^{k-1}  (\mathbf{1}_0, {\tiny \yng(1)}_{\,1})^{N-k} \nonumber \\
  & \sim  V^{U(N-1)}_{+} W_{\alpha}^{k-1}  Q^{N-k},
\end{align}
where the color indices of the gaugino and matter fields are contracted by an epsilon tensor of the unbroken $U(N-1)$ subgroup. From the quantum numbers of these dressed states, we can identify them with $V_{d\pm}$.

These dressed operators $V_{d \pm}$ describe the Coulomb-Higgs mixed branch where the baryonic operators are also turned on in addition to the adjoint scalar in the vector multiplet. The second construction of the dressed monopoles naturally appear in the SCI expansion since the bare monopoles $V^{U(N-1)}_{\pm}$ are associated with the GNO charges $(1,0\cdots,0)$ and $(0,\cdots,-1)$. From the viewpoint of the moduli space, this construction is not fully satisfactory because the Coulomb flat direction spanned by $V^{U(N-1)}_{\pm}$ becomes massive due to the $U(1)_{CB}$ CS term.

\begin{table}[H]\caption{The 3d $\mathcal{N}=2$ $U(N)_{k,k-N}$ gauge theory with $F({\tiny \protect\yng(1)}_{\, 1}+ {\tiny \overline{\protect\yng(1)}}_{\,-1})$} 
\begin{center}
\scalebox{0.95}{
  \begin{tabular}{|c||c|c|c|c|c|c| } \hline
  &$U(N)_{k,k-N}$&$SU(F)$&$SU(F)$&$U(1)_A$&$U(1)_T$&$U(1)_R$ \\ \hline
$Q$& ${\tiny \yng(1)}_{\, 1}$&$\tiny \yng(1)$&1&$1$&0&$0$ \\  
$\tilde{Q}$& ${\tiny \overline{\yng(1)}}_{\, -1}$&1&$\tiny \yng(1)$&1&0&0  \\ \hline
$M:=Q\tilde{Q}$&1&${\tiny \yng(1)}$&$\tiny \yng(1)$&$2$&0&$0$ \\   \hline
 $V_{\pm}$&$U(1)_P:$ $\pm(N-k)$&1&1&$-F$&$\pm 1$&$F-N+k$ \\
$V_{d-}:=V_- Q^{N-k}$&1 &$\left[ N-k \right]$&1&$N-F-k$&$-1$&$F-N+k$ \\
 $V_{d+}:=V_+ \tilde{Q}^{N-k}$&1 &1&$\left[ N-k \right]$&$N-F-k$&$+1$&$F-N+k$ \\  \hline
  \end{tabular}}
  \end{center}\label{GKelenm1}
\end{table}

For $n=-1$, the magnetic side becomes a 3d $\mathcal{N}=2$ $\stackrel{+1~~~~~~~~~~~~~~~~~~~~~~~~~~}{\protect\wick{1}{<1 U(1)_{0} \times >1 U(F+k-N)_{-k,F-N} }}$ gauge theory with $F$ dual flavors and a meson singlet $M$. Since the CS level for the topological $U(1)$ gauge group vanishes and the matter fields are not charged under the topological symmetry, we can integrate over the $U(1)_0$ vector multiplet. Due to the BF interaction between $U(1)_0$ and $U(F+k-N)_{-k,F-N}$, the overall $U(1) \subset U(F+k-N)_{-k,F-N}$ is also turned off through the integration of the topological $U(1)$ vector multiplet. As a result, the dual description becomes a 3d $\mathcal{N}=2$ $SU(F+k-N)_{-k}$ gauge theory with $F$ dual flavors and a meson singlet (see Table \ref{GKmagnm1}). The theory has a tree-level superpotential $W_{mag}=Mq \tilde{q}$ as before. This duality can be also obtained from the $SU(N)_k$ duality proposed in \cite{Aharony:2013dha, Aharony:2014uya, Park:2013wta} by swapping the electric and magnetic descriptions. As studied in \cite{Aharony:2013dha, Aharony:2014uya}, there is no Coulomb branch of the moduli space. Since the gauge group is special unitary, there is a baryonic direction of the moduli space
\begin{align}
V_{d-} \sim q^{F+k-N},~~~~~~~V_{d+} \sim \tilde{q}^{F+k-N}.
\end{align}
From the quantum numbers of these magnetic baryons, we find that the magnetic baryons are mapped to the dressed Coulomb branch under the duality.  

\begin{table}[H]\caption{The magnetic $SU(F+k-N)_{-k} $ gauge theory dual to Table \ref{GKelenm1}} 
\begin{center}
\scalebox{0.97}{
  \begin{tabular}{|c||c|c|c|c|c|c| } \hline
  &$SU(F+k-N)_{-k} $&$SU(F)$&$SU(F)$&$U(1)_A$&$U(1)_T$&$U(1)_R$ \\ \hline
$q$& ${\tiny \yng(1)}_{\, 1}$&$\tiny \overline{\yng(1)}$&1&$-1$&$-\frac{1}{F+k-N}$&$1$ \\  
$\tilde{q}$& ${\tiny \overline{\yng(1)}}_{\, -1}$&1&$\tiny \overline{\yng(1)}$&$-1$&$+\frac{1}{F+k-N}$&1  \\ 
$M$&1&${\tiny \yng(1)}$&$\tiny \yng(1)$&$2$&0&$0$ \\   \hline
 $V_{d-} \sim q^{F+k-N}$&1&$\left[ N-k \right]$&1&$N-F-k$&1&$F+k-N$ \\
 $V_{d+} \sim \tilde{q}^{F+k-N}$&1&1&$\left[ N-k \right]$&$N-F-k$&1&$F+k-N$ \\ \hline
  \end{tabular}}
  \end{center}\label{GKmagnm1}
\end{table}

\section{$n=-2$: $U(N)_{k,k-2N}$ Generalized GK duality}
For $n=-2$, the electric description becomes a 3d $\mathcal{N}=2$ $U(N)_{k,k-2N}$ gauge theory with $F$ fundamental flavors. For some combinations of $(N,k)$, the theory allows a Coulomb moduli space. We here show two cases. 

The first example is the case with $k=2N$ where the abelian CS level is vanishing. The Coulomb branch associated to the overall $U(1)$ vector multiplet is parametrized by the vacuum expectation value of the adjoint scalar
\begin{align}
\braket{\phi_{adj.}} =\mathrm{diag.} (v,\cdots,v).
\end{align}
The corresponding monopole operator $X_{\pm}^{k=2N}$ is constructed by dualizing the $U(1) \subset U(N)$ vector multiplet into a chiral superfield. Since the abelian CS term is zero, the flat direction $X_{\pm}^{k=2N}$ becomes exactly massless. The subscript of $X_{\pm}^{k=2N}$ means that the Coulomb branch is split into two regions with positive or negative vevs. Their quantum numbers are computed as in Table \ref{GKelenm2}. Note that the bare monopole is gauge-invariant since there is only a single $U(1)$ gauge group and no mixed CS term is generated.

The second example is the case with $k=2$ where the gauge group becomes $U(N)_{2,2-2N}$. In this case, we have to consider the different Coulomb branch (monopole operator) $X_{\pm}^{k=2}$ whose insertion leads to the gauge symmetry breaking
\begin{align}
U(N)_{2,2-2N} & \rightarrow  \stackrel{-2~~~~~~~~~~~~~~~\,~~~~~~~~}{\protect\wick{1}{<1 U(1)_{0} \times >1 U(N-1)_{2,2-2(N-1)} }} \\
   {\tiny \yng(1)}_{\,1}  & \rightarrow  ({\tiny \yng(1)}_{\,1}  , 1_{\,0})+ (1_{\,0}, {\tiny \yng(1)}_{\,1}  )\\ 
{\tiny \overline{\yng(1)}}_{\,-1}  & \rightarrow  ({\tiny \overline{\yng(1)}}_{\,-1} , 1_{\,0}) +(1_{\,0},{\tiny \overline{\yng(1)}}_{\,-1} ).
\end{align}
The associated monopole is defined for the $U(1)_0$ subgroup. This Coulomb branch is realized by introducing a vacuum expectation value for the adjoint scalar in the $U(N)$ vector multiplet as $\braket{\phi_{adj.}} =\mathrm{diag.} (\pm v,0,\cdots,0)$.
Since the CS level for the unbroken $U(1)$ subgroup is zero, the Coulomb branch $X_{\pm}^{k=2}$ is exactly flat. Notice that the CS term behaves as a topological mass term for the vector multiplet and that the classical Coulomb flat direction with a non-zero CS term is eliminated from the quantum moduli space. 
Due to the mixed CS term, the bare monopoles $X_{\pm}^{k=2}$ obtain non-zero $U(1) \subset U(N-1)$ charges \cite{Intriligator:2013lca}. Therefore, we need to define the dressed operators
\begin{align}
X_{+d}^{k=2}&:= X_{+}^{k=2} ((1_{\,0},{\tiny \overline{\yng(1)}}_{\,-1} ))^{2(N-1)} \sim X_{+}^{k=2}  \tilde{Q}^{2(N-1)} \\
X_{-d}^{k=2} &:=X_{-}^{k=2}  ((1_{\,0}, {\tiny \yng(1)}_{\,1}  ))^{2(N-1)} \sim X_{-}^{k=2} Q^{2(N-1)}, 
\end{align}
where the color indices of $Q^{2(N-1)}$ and $\tilde{Q}^{2(N-1)}$ are contracted by two epsilon tensors of the $U(N-1)$ gauge group. The quantum numbers of these operators are summarized in Table \ref{GKelenm2}. In the table, the symbol $[N-1]$ means that the $N-1$ flavor indices are anti-symmetrized.

\begin{table}[H]\caption{The 3d $\mathcal{N}=2$ $U(N)_{k,k-2N}$ gauge theory with $F(\,{\tiny \protect\yng(1)}_{\, 1}+ {\tiny \overline{\protect\yng(1)}}_{\,-1})$} 
\begin{center}
\scalebox{1}{
  \begin{tabular}{|c||c|c|c|c|c|c| } \hline
  &$U(N)_{k,k-2N}$&$SU(F)$&$SU(F)$&$U(1)_A$&$U(1)_T$&$U(1)_R$ \\ \hline
$Q$& ${\tiny \yng(1)}_{\, 1}$&$\tiny \yng(1)$&1&$1$&0&$0$ \\  
$\tilde{Q}$& ${\tiny \overline{\yng(1)}}_{\, -1}$&1&$\tiny \yng(1)$&1&0&0  \\ \hline
$M:=Q\tilde{Q}$&1&${\tiny \yng(1)}$&$\tiny \yng(1)$&$2$&0&$0$ \\   \hline
$X_{\pm}^{k=2N}$&1&1&1&$-NF$&$\pm 1$&$NF$  \\
$X_{+ d}^{k=2}$&1&1&$[N-1]^2$&$2N-F-2$&$ + 1$&$F-N+1$  \\ 
$X_{- d}^{k=2}$&1&$[N-1]^2$&1&$2N-F-2$&$ -1$&$F-N+1$  \\ \hline 
  \end{tabular}}
  \end{center}\label{GKelenm2}
\end{table}

The magnetic side becomes a 3d $\mathcal{N}=2$ $\stackrel{+1~~~~~~~~~~~~~~~~~~~~~~~~~~}{\protect\wick{1}{<1 U(1)_{-1} \times >1 U(F+k-N)_{-k,F-N} }}$ gauge theory with $F$ dual flavors and a meson singlet $M$. The theory has a tree-level superpotential $W_{mag}=Mq \tilde{q}$. Since the CS term for the topological $U(1)$ gauge symmetry is $-1$, we can integrate over it as follows: By taking a linear combination of the two $U(1)$ gauge symmetries, we define a new topological $U(1)$ symmetry 
\begin{align}
B_\mu = A_\mu^{top} -(F+k-N)A_\mu^{U(1) \subset U(F+k-N)}.
\end{align}
In this new basis of the abelian gauge fields $\left\{ B_\mu, A_\mu^{U(1) \subset U(F+k-N)} \right\}$, the product gauge group is recast into $U(1)^{B}_{-1} \times U(F+k-N)_{-k,2F+k-2N}$ without a mixed CS term. The chiral multiplets are not charged under the topological $U(1)^B_{-1}$ symmetry. Hence, we can integrate out the $U(1)^{B}_{-1}$ vector multiplet. The resultant gauge group simply becomes $U(F+k-N)_{-k,2F+k-2N}$ and the matter content is unchanged (see Table \ref{GKmagnm2}). This case corresponds to the duality reported in \cite{Choi:2018ohn}. As a consistency check of the duality, let us study the Coulomb moduli space for the two cases with $k=2N$ and $k=2$.

For $k=2N$, the magnetic gauge group becomes $U(F+N)_{-2N, 2F}$. Along the magnetic Coulomb branch, which is denoted by $\tilde{X}_{\pm}^{k=2N}$, the gauge group is spontaneously broken as
\begin{align}
U(F+N)_{-2N,2F} & \rightarrow \stackrel{+2~~~~~~~~~~~~~\,~~~~~~~~}{\protect\wick{1}{<1 U(N)_{-2N,0} \times >1 U(F)_{-2N,-2N+2F} }} \\
  {\tiny \yng(1)}_{\,1}  & \rightarrow  (  {\tiny \yng(1)}_{\,1}, 1_{\,0})+(1_{\,0} ,  {\tiny \yng(1)}_{\,1})  \\
  {\tiny \overline{\yng(1)}}_{\,-1}  & \rightarrow  (  {\tiny \overline{\yng(1)}}_{\,-1}, 1_{\,0})+(1_{\,0},   {\tiny \overline{\yng(1)}}_{\,-1}),
\end{align}
where the moduli space is spanned by the vacuum expectation values of the adjoint scalar in the $U(F+N)$ vector multiplet 
\begin{align}
\braket{\tilde{\phi}_{adj.}} =\mathrm{diag.} ( \overbrace{v,\cdots,v}^{N}, \overbrace{0,\cdots,0}^F).
\end{align}
Depending on the sign of $v$, there are two Coulomb branch coordinates $\tilde{X}_{\pm}^{k=2N}$. Since the $U(1) \subset U(N)$ CS level is vanishing, the corresponding Coulomb branch is quantum-mechanically massless. 
Due to the mixed CS term between the $U(N)$ and $U(F)$ subgroups, the bare monopoles $\tilde{X}_{\pm}^{k=2N}$ are charged under the $U(1) \subset U(F)$ symmetry \cite{Intriligator:2013lca, Csaki:2014cwa, Amariti:2015kha, Nii:2018bgf}. As a result, the gauge-invariant composites are defined by
\begin{align}
\tilde{X}^{k=2N}_{+} &:= \tilde{X}_+^{k=2N} (1_{\,0} ,  {\tiny \yng(1)}_{\,1})^{2NF}  \sim  \tilde{X}_+  Q^{2NF} \\
\tilde{X}^{k=2N}_{-} &:= \tilde{X}_-^{k=2N} (1_{\,0},   {\tiny \overline{\yng(1)}}_{\,-1})^{2NF} \sim  \tilde{X}_-  \tilde{Q}^{2NF},
\end{align}
where the color indices of the matter multiplets are contracted by $2N$ epsilon tensors of the unbroken $U(F)$ gauge group. From the quantum numbers of these dressed monopoles, we obtain the identification $X_{\pm}^{k=2N} \sim \tilde{X}_{\pm}^{k=2N}$

Let us move on to the second case with $k=2$, where the magnetic gauge group becomes $U(F+2-N)_{-2,2F+2-2N}$. In this case, the Coulomb branch $\tilde{X}_{\pm}^{k=2}$ corresponds to the gauge symmetry breaking
\begin{align}
U(F+2-N)_{-2,-2+2(F+2-N)} & \rightarrow  \stackrel{+2~~~~~~~~~~~~~~~~~~~~~~~~~~~\,~~~~~~~~}{\protect\wick{1}{<1 U(1)_{0} \times >1 U(F+1-N)_{-2,-2+2(F+1-N)} }} \\
   {\tiny \yng(1)}_{\,1}  & \rightarrow  ({\tiny \yng(1)}_{\,1}  , 1_{\,0})+ (1_{\,0}, {\tiny \yng(1)}_{\,1}  )\\ 
{\tiny \overline{\yng(1)}}_{\,-1}  & \rightarrow  ({\tiny \overline{\yng(1)}}_{\,-1} , 1_{\,0}) +(1_{\,0},{\tiny \overline{\yng(1)}}_{\,-1} ) .
\end{align}
The Coulomb branch is associated with the flat direction of the $U(1)_0$ vector superfield and its CS level is correctly zero as it should be. Due to the level-2 mixed CS term, the bare Coulomb branch $\tilde{X}_{\pm}^{k=2}$ is not gauge-invariant. 
In order to cancel the $U(1) \subset U(F+1-N)$ charges of the bare monopoles $\tilde{X}_{\pm}^{k=2}$, we need to define the dressed operators \cite{Intriligator:2013lca, Csaki:2014cwa, Amariti:2015kha, Nii:2018bgf} 
\begin{align}
\tilde{X}_{+d}^{k=2} &:= \tilde{X}_+^{k=2}  ((1_{\,0}, {\tiny \yng(1)}_{\,1}  ))^{2(F+1-N)} \sim \tilde{X}_+ q^{2(F+1-N)}  \\
 \tilde{X}_{-d}^{k=2} &:= \tilde{X}_-^{k=2} ((1_{\,0},{\tiny \overline{\yng(1)}}_{\,-1} ))^{2(F+1-N)}  \sim  \tilde{X}_-  \tilde{q}^{2(F+1-N)},
\end{align}
which are identified with the electric dressed operators $X_{\pm d}^{k=2}$. These operators parametrize the Coulomb-Higgs mixed branch of the moduli space. 

\begin{table}[H]\caption{The magnetic $U(F+k-N)_{-k,2F+k-2N} $ gauge theory dual to Table \ref{GKelenm2}} 
\begin{center}
\scalebox{0.95}{
  \begin{tabular}{|c||c|c|c|c|c|c| } \hline
  &$U(F+k-N)_{-k,2F+k-2N} $&$SU(F)$&$SU(F)$&$U(1)_A$&$U(1)_T$&$U(1)_R$ \\ \hline
$q$& ${\tiny \yng(1)}_{\, 1}$&$\tiny \overline{\yng(1)}$&1&$-1$&0&$1$ \\  
$\tilde{q}$& ${\tiny \overline{\yng(1)}}_{\, -1}$&1&$\tiny \overline{\yng(1)}$&$-1$&0&1  \\ 
$M$&1&${\tiny \yng(1)}$&$\tiny \yng(1)$&$2$&0&$0$ \\   \hline
$X_{\pm}^{k=2N}$&1&1&1&$-NF$&$\pm 1$&$NF$  \\
$X_{+ d}^{k=2}$&1&1&$[N-1]^2$&$2N-F-2$&$ \pm 1$&$F-N+1$  \\ 
$X_{- d}^{k=2}$&1&$[N-1]^2$&1&$2N-F-2$&$ \pm 1$&$F-N+1$  \\ \hline 
  \end{tabular}}
  \end{center}\label{GKmagnm2}
\end{table}

\section{Examples}
In this section, we investigate the proposed duality especially for the $U(2)$ and $U(1)$ gauge groups without restricting the value of $n$. As a non-trivial test of the duality, we will also compute the superconformal indices \cite{Bhattacharya:2008bja, Kim:2009wb, Imamura:2011su, Kapustin:2011jm} by employing the localization technique \cite{Pestun:2007rz, Kapustin:2009kz, Hama:2010av}. We will observe a nice agreement of the indices under the duality. By carefully studying the indices, we can find a leading monopole operator and its mapping under the duality. For some values of $n$, the theory possesses the Coulomb branch in the moduli space and the corresponding monopole operator is presented.

\subsection{$U(2)_{2,2+2n}$ with a single flavor}
The first example is a 3d $\mathcal{N}=2$ $U(2)_{2,2+2n}$ gauge theory with a single flavor, which corresponds to the case with $N=k=2,$ and $F=1$. The quantum numbers of the elementary fields are summarized in Table \ref{GGKU(2)F=1ele}. We here listed a generic r-charge by mixing the r-symmetry with the axial $U(1)_A$ symmetry in Table \ref{GGKele}. 

The dual description becomes a 3d $\mathcal{N}=2$ $\stackrel{+1~~~~~~}{\protect\wick{1}{<1 U(1)_{n+1} \times >1 U(1)_{-1} }}$ gauge theory with a dual flavor and a gauge-singlet meson $M$. The theory includes a tree-level superpotential $W_{mag}=Mq \tilde{q}$. Table \ref{GGKU(2)F=1mag} summarizes the quantum numbers of the dual fields. In what follows, we will give a detailed analysis for small $n$.

\begin{table}[H]\caption{The 3d $\mathcal{N}=2$ $U(2)_{2,2+2n}$ gauge theory with ${\tiny \protect\yng(1)}_{\, 1}+ {\tiny \overline{\protect\yng(1)}}_{\,-1}$} 
\begin{center}
\scalebox{1}{
  \begin{tabular}{|c||c|c|c|c| } \hline
  &$U(2)_{2,2+2n}$&$U(1)_A$&$U(1)_T$&$U(1)_R$ \\ \hline
$Q$& ${\tiny \yng(1)}_{\, 1}$&$1$&0&$r$ \\  
$\tilde{Q}$& ${\tiny \overline{\yng(1)}}_{\, -1}$&1&0&$r$  \\ \hline
$M:= Q\tilde{Q}$&$\mathbf{1}$&2&0&$2r$  \\ \hline
  \end{tabular}}
  \end{center}\label{GGKU(2)F=1ele}
\end{table}

\begin{table}[H]\caption{The magnetic $\stackrel{+1~~~~~~}{\protect\wick{1}{<1 U(1)_{n+1} \times >1 U(1)_{-1} }}$ gauge theory dual to Table \ref{GGKU(2)F=1ele}} 
\begin{center}
\scalebox{1}{
  \begin{tabular}{|c||c|c|c|c| } \hline
  &$\stackrel{+1~~~~~~}{\protect\wick{1}{<1 U(1)_{n+1} \times >1 U(1)_{-1} }}$&$U(1)_A$&$U(1)_T$&$U(1)_R$ \\ \hline
$q$& $(0,1)$&$-1$&0&$1-r$ \\  
$\tilde{q}$& $(0,-1)$&$-1$&0&$1-r$  \\ 
$M$&$\mathbf{1}$&2&0&$2r$  \\ \hline
  \end{tabular}}
  \end{center}\label{GGKU(2)F=1mag}
\end{table}

\subsubsection*{The case with $n=1$}
For $n=1$, the theory has no Coulomb moduli space since the the bare CS levels of the $U(2)_{2,4}$ gauge group cannot be canceled in any directions of the classical Coulomb branch. This situation is completely the same as the magnetic side where the gauge group is $U(1)_{2} \times U(1)_{-1}$ and the CS levels cannot be canceled by the vector-like matter. 
As a result, the theory only has the Higgs branch which is described by $M:=Q \tilde{Q}$. We can see a nice agreement of the superconformal indices between the electric and magnetic descriptions
\begin{align}
I_{n=1} &:=1+t^2 \sqrt{x}+t^4 x+t^6 x^{3/2}+\left(t^8-2\right) x^2+t^{10} x^{5/2}+t^{12} x^3+\left(t^{14}+\frac{1}{t^2}\right) x^{7/2} \nonumber \\
&\qquad \qquad +\left(t^{16}-3\right) x^4+\left(t^{18}-2 t^2\right) x^{9/2}+t^{20} x^5+\cdots.
\end{align}
We set the r-charge to be $r=\frac{1}{4}$ for simplicity although we can see this agreement for other choices of $r$. $t$ is a fugacity parameter for the $U(1)_A$ symmetry. The second term $t^2 \sqrt{x}$ is regarded as the meson $M:=Q\tilde{Q}$. The fermion contribution $-2 x^2$ is interpreted as $\psi_Q Q+\psi_{\tilde{Q}} \tilde{Q}$, where $\psi$ denotes the fermion partner of the squark.

\subsubsection*{The case with $n=0$}
For $n=0$, the electric gauge group becomes $U(2)_{2,2}$. There is no Coulomb branch of the moduli space for the same reason as the $n=1$ case. Therefore, the superconformal indices are agin simplified a lot. We observed that the electric and magnetic indices beautifully agree with each other and the result is expanded as    
\begin{align}
I_{n=0}&:=1+t^2 \sqrt{x}+t^4 x+t^6 x^{3/2}+\left(t^8-2\right) x^2+t^{10} x^{5/2}+t^{12} x^3+\left(t^{14}+\frac{1}{t^2}\right) x^{7/2} \nonumber \\
&\qquad \qquad  +\frac{2 x^{15/4}}{t}+\left(t^{16}-3\right) x^4+\left(t^{18}-2 t^2\right) x^{9/2}+t^{20} x^5+\cdots,
\end{align}
where $t$ is a fugacity parameter for the global $U(1)_A$ symmetry and the r-charge is set to be $r=\frac{1}{4}$. The second term $t^2 \sqrt{x}$ is identified with the meson $M:=Q \tilde{Q}$ and the higher-order terms can be interpreted in the same way as the $n=1$ case. In our choice of the r-charge, the only difference up to $O(x^5)$ appears as $\frac{2 x^{15/4}}{t}$. We will claim that this term can be understood as the dressed monopole operators which are not part of the moduli space: On the electric side, the insertion of the monopole $X_{\pm}$ leads to the gauge symmetry breaking
\begin{align}
U(2)_{2,2} &\rightarrow U(1)_2 \times U(1)_2   \\
 {\tiny \yng(1)}_{\,1}  & \rightarrow  (+1, 0)+ (0, +1)\\ 
{\tiny \overline{\yng(1)}}_{\,-1}  & \rightarrow  (-1, 0) +(0,-1) \\
\mathbf{adj.}& \rightarrow  (0,0)+(0,0)+(+1,-1)+(-1,+1),
\end{align}
where $\mathbf{adj.}$ denotes a gaugino field. 
Since $n$ is vanishing, there is no mixed CS term between the two $U(1)_2$ subgroups. The monopole here is associated with the first $U(1)_2$ factor. Due to the CS term indicated above, the bare operator $X_{\pm}$ obtains a non-zero charge under the first $U(2)_2$ subgroup. We can see that the dressed composites
\begin{align}
X_{+}^{dressed} &:=X_{+} (+1,0) (0,1) (+1,-1) \sim X_{+} Q \psi_{\tilde{Q}} W_\alpha \\
X_{-}^{dressed} &:=X_{-} (-1,0) (0,-1) (-1,+1) \sim X_{-} \tilde{Q} \psi_{Q} W_\alpha
\end{align}
become gauge-invariant and explain the states with the quantum numbers of $\frac{2 x^{15/4}}{t}$. Note that under the monopole background, the spin of the components charged under the first $U(1)_2$ subgroup are transmuted \cite{Polyakov:1988md, Dimofte:2011py, Aharony:2015pla}. Therefore, the above combinations contribute as bosons. 

On the magnetic side, we need to consider two types of monopole operators in the $U(1)_1 \times U(1)_{-1}$ gauge group. For each $U(1)$, we can define a pair of monopole operators.  We denote the monopoles associated with $U(1)_1$ by $\tilde{\mathcal{U}}_{\pm}$ and the monopoles associated with $U(1)_{-1}$ by $\tilde{X}_{\pm}$. Due to the (mixed) CS terms, these operators are not gauge-invariant. We find that the (magnetic) dressed monopoles are defined by 
\begin{align}
\tilde{\mathcal{U}}_{+} \tilde{X}_{-} q^2,~~~~\tilde{\mathcal{U}}_{-} \tilde{X}_{+}  \tilde{q}^2.
\end{align}
From the symmetry argument, these are identified with the electric dressed operators $X_{\pm}^{dressed}$ and expressed as $\frac{2 x^{15/4}}{t}$ in the superconformal indices. This observation confirms the validity of our duality proposal.


\subsubsection*{The case with $n=-1$}
For $n=-1$, the electric theory allows a pair of the monopole operators $V_{\pm}$ which are associated with the overall $U(1) \subset U(2)_{2,2+2n}$ Coulomb branch. Notice that the CS term for the abelian factor vanishes when $n=-1$ and the corresponding Coulomb branch can be exactly massless. Since there is a single abelian gauge group, the bare operators $V_{\pm}$ are gauge-invariant. 

On the magnetic side, the CS term for the topological (gauged) $U(1)$ symmetry becomes zero. Therefore, there are two monopole operators $\tilde{\mathcal{U}}_{\pm}$ describing the associated Coulomb branch. Due to the mixed CS term in $ \stackrel{+1~~~~~\,~}{\protect\wick{1}{<1 U(1)_{n+1} \times >1 U(1)_{-1} }} |_{n=-1}$, the bare operators $\tilde{\mathcal{U}}_{\pm}$ obtain a non-zero $U(1)_{-1}$ charge. Therefore, we need to define the dressed monopoles
\begin{align}
\tilde{\mathcal{U}}_{+}q,~~~~\tilde{\mathcal{U}}_{-} \tilde{q}. 
\end{align}
From the symmetry argument, these are identified with the electric monopole operators $V_{\pm}$. We can see these contributions in the superconformal indices as follows: The electric and magnetic indices are expanded as  
\begin{align}
I_{n=-1}&:=1+t^2 \sqrt{x}+\frac{2 x^{3/4}}{t}+t^4 x+\left(t^6+\frac{2}{t^2}\right) x^{3/2}+\left(t^8-2\right) x^2+\frac{2 x^{9/4}}{t^3}+t^{10} x^{5/2} \nonumber \\
&\qquad +\left(t^{12}+\frac{2}{t^4}\right) x^3+2 t x^{13/4}+\left(t^{14}+\frac{1}{t^2}\right) x^{7/2}+\frac{2 x^{15/4}}{t^5}+\left(t^{16}-3\right) x^4  \nonumber \\
&\qquad \qquad  +\left(t^{18}+\frac{2}{t^6}-2 t^2\right) x^{9/2}-\frac{4 x^{19/4}}{t}+t^{20} x^5+\cdots, 
\end{align}
where the r-charges of $Q$ and $\tilde{Q}$ are set to be $r=\frac{1}{4}$ for simplicity. For different values of $r$, one can easily check the agreement of the electric and magnetic indices. The parameter $t$ denotes the fugacity for the $U(1)_A$ symmetry. The second term $t^2 \sqrt{x}$ is a meson $M:=Q\tilde{Q}$ contribution. The bare monopoles $V_{\pm}$ (or $\tilde{\mathcal{U}}_{+}q$ and $\tilde{\mathcal{U}}_{-} \tilde{q}$ from the magnetic viewpoint) appear as $\frac{2 x^{3/4}}{t}$, which is consistent with the global charges of $V_{\pm}$. The higher-order terms are interpreted as symmetric products of these operators and the fermion contributions. 


\subsubsection*{The case with $n=-2$}
For $n=-2$, there is a pair of the Coulomb branch coordinates. These are parametrized by the bare monopoles $X_{\pm}^{bare}$ which induce the gauge symmetry breaking
\begin{align}
U(2)_{2,2+2n} &\rightarrow  \stackrel{+n~~~~~~~~}{\protect\wick{1}{<1 U(1)_{2+n} \times >1 U(1)_{2+n} }}\\
  {\tiny \yng(1)}_{\,1}  & \rightarrow  (+1, 0)+ (0, +1)\\ 
{\tiny \overline{\yng(1)}}_{\,-1}  & \rightarrow  (-1, 0) +(0,-1),
\end{align}
where the Coulomb branch is associated with the first $U(1)_{2+n}$ subgroup and the CS level correctly vanishes for $n=-2$. Notice that the mixed CS term is also generated between the two $U(1)_{2+n}$ gauge groups. Therefore, the bare monopole operators $X_{\pm}^{bare}$ are charged under the second $U(1)_{2+n}$ subgroup. The gauge-invariant operators are constructed by dressing the bare operator with the massless chiral multiplets:
\begin{align}
X_{+d} &:= X_{+}^{bare} (0,-1)^2 \sim X_{+}^{bare} \tilde{Q}^2 \\
X_{-d} &:= X_{-}^{bare} (0,+1)^2 \sim X_{+}^{bare} Q^2 .
\end{align}

On the magnetic side, the gauge group becomes $ \stackrel{+1~~~~~~~~}{\protect\wick{1}{<1 U(1)_{-1} \times >1 U(1)_{-1} }}$ for $n=-2$. Then, one might consider that there is no magnetic Coulomb branch since the CS terms cannot be canceled along the (classical) Coulomb moduli space. However, by taking a linear combination of the two gauged $U(1)$ symmetries 
\begin{align}
A_{\mu}^{top, new}&:= A_{\mu}^{top. U(1)} - A^{U(1)}_{\mu},
\end{align}
the CS term is only introduced for the new topological $U(1)$ symmetry. The resulting gauge group becomes $U(1)_{-1} \times U(1)_0$ with no mixed CS term. Therefore, there is a pair of monopole operators $\tilde{X}_{\pm}$ associated with the latter $U(1)_0$ gauge dynamics.  Since $\tilde{X}_{\pm}$ are gauge-invariant, these are directly identified with $X_{\pm d}$.

As a test of the duality, we can compare the electric and magnetic superconformal indices which capture the dressed Coulomb branch coordinates studied above. The indices are expanded as
\begin{align}
I_{n=-2} &:=1+2 t x^{1/4}+3 t^2 x^{1/2}+4 t^3 x^{3/4}+5 t^4 x+6 t^5 x^{5/4}+7 t^6 x^{3/2}+\left(8 t^7-\frac{2}{t}\right) x^{7/4} \nonumber  \\
&+\left(9 t^8-4\right) x^2+\left(10 t^9-4 t\right) x^{9/4}+\left(11 t^{10}-4 t^2\right) x^{5/2}+\left(12 t^{11}-4 t^3\right) x^{11/4}  \nonumber \\
&+\left(13 t^{12}-4 t^4\right) x^3+\left(12 t^{13}-4 t^5\right) x^{13/4}+\left(13 t^{14}-4 t^6+\frac{1}{t^2}\right) x^{7/2} \nonumber \\
&+\left(12 t^{15}-4 t^7\right) x^{15/4}+\left(13 t^{16}-4 t^8-5\right) x^4+\left(12 t^{17}-4 t^9-8 t\right) x^{17/4} \nonumber \\
&+\left(13 t^{18}-4 t^{10}-8 t^2\right) x^{9/2}+\left(12 t^{19}-2 t^{11}-8 t^3\right) x^{19/4}+\left(13 t^{20}-2 t^{12}-8 t^4\right) x^5+\cdots, 
\end{align}
where the r-charge is set to $r=\frac{1}{4}$ and $t$ is a fugacity parameter for the $U(1)_A$ symmetry. The dressed monopoles $X_{\pm d}$ (or the bare monopole $\tilde{X}_{\pm}$ on the magnetic side) are expressed as $2 t x^{1/4}$. The third term $3 t^2 x^{1/2}$ consists of the three contributions $M+X_{+ d}^2+X_{- d}^2$.

\subsubsection*{The case with $n=-3$}
For $n=-3$, the lower-order terms of the superconformal indices are again simplified a lot since there is no Coulomb moduli space. We computed the electric and magnetic indices and observed a nice agreement. The result is given by
\begin{align}
I_{n=-3}&:=1+t^2 \sqrt{x}+t^4 x+t^6 x^{3/2}+\left(t^8-2\right) x^2+t^{10} x^{5/2}+t^{12} x^3+2 t x^{13/4}+\left(t^{14}+\frac{1}{t^2}\right) x^{7/2}  \nonumber \\
&\qquad \qquad  +\left(t^{16}-3\right) x^4+\left(t^{18}-2 t^2\right) x^{9/2}-\frac{2 x^{19/4}}{t}+t^{20} x^5+\cdots,
\end{align}
where the r-charge is set to be $r=\frac{1}{4}$ for simplicity. $t$ denotes a fugacity parameter for the $U(1)_A$ symmetry. The second term represents a meson operator $M:=Q \tilde{Q}$. The expansion of the index is very similar to the one of the $n=1$ case. The difference first appears as $2 t x^{13/4}$ and we will give its operator interpretation as a leading monopole operator.  

On the electric side, this term can be regarded as the dressed monopole states
\begin{align}
X_{+} (-1,0) (0,-1)^3  \sim  X_{+} \psi_{Q} \tilde{Q}^3  \label{fnonzero1}\\
X_{-} (+1,0) (0,+1)^3  \sim  X_{-} \psi_{\tilde{Q}} Q^3,   \label{fnonzero2}
\end{align}
where the insertion of the bare monopole $X_{\pm}$ induces the gauge symmetry breaking
\begin{align}
U(2)_{2,-4} &\rightarrow  \stackrel{-3~~~~~~~~}{\protect\wick{1}{<1 U(1)_{-1} \times >1 U(1)_{-1} }}\\
  {\tiny \yng(1)}_{\,1}  & \rightarrow  (+1, 0)+ (0, +1)\\ 
{\tiny \overline{\yng(1)}}_{\,-1}  & \rightarrow  (-1, 0) +(0,-1) \\
\mathbf{adj.}& \rightarrow  (0,0)+(0,0)+(+1,-1)+(-1,+1).
\end{align}
The bare operators $X_{\pm}$ are charged under the $U(1)_{-1} \times U(1)_{-1}$ subgroup due to the (mixed) CS terms. Although there are many gauge-invariant combinations which might contribute to the index at $O(x^{13/4})$, they are precisely canceled out since the following combinations
\begin{align}
W_{\alpha} \tilde{Q} = (-1,1) (0,-1),~~~~~~\tilde{Q} =(-1,0)
\end{align}
have the same quantum numbers except for their spin-statistics. The similar argument can hold also for the fundamental quark $Q$ and then there are enormous cancellations in the superconformal indices. The leading non-zero contribution comes from \eqref{fnonzero1} and \eqref{fnonzero2}.

On the magnetic side, the gauge group becomes $U(1)_{-2} \times U(1)_{-1}$ for $n=-3$. For the first $U(1)_{-2}$ subgroup, we denote the monopole operator by $\tilde{\mathcal{U}}_{\pm}$. For the second $U(1)_{-1}$ subgroup, we denote the bare monopole by $\tilde{X}_{\pm}$. These bare operators are not gauge-invariant due to the (mixed) CS terms. Therefore, we need to construct the dressed monopoles by combining them with the chiral multiplets
\begin{align}
\tilde{\mathcal{U}}_+ \tilde{X}_+^2 \tilde{q},~~~~~~\tilde{\mathcal{U}}_{-} \tilde{X}_-^2 q. 
\end{align}
From the quantum numbers of these dressed states, these composites appear as $2 t x^{13/4}$ in the superconformal indices. This observation supports the validity of the duality proposal. 


\subsection{$U(2)_{1,1+2n}$ with two flavors}
Next, we consider the 3d $\mathcal{N}=2$ $U(2)_{1,1+2n}$ gauge theory with two flavors in a fundamental representation. The quantum numbers of the electric fields are summarized in Table \ref{GGKU(2)F=2ele}. The Higgs branch is described by the meson composite $M:=Q \tilde{Q}$.
We here focus on the Coulomb branch of the electric theory. The bare Coulomb branch induces the gauge symmetry breaking $U(2)_{1,1+2n} \rightarrow U(1)_{1+n} \times U(1)'_{1+n}$. There is also a mixed CS term with level-$n$ between the $U(1)$ and $U(1)'$ subgroup. The fact that the non-zero CS terms exist for the $U(1)$ subgroups means that there is no flat direction from the vector multiplet except for $n=-1$. This situation is different from the previous example where the CS term for the overall $U(1) \subset U(2)$ gauge group could be zero (for $n=-2$) while it is now impossible for the $U(2)_{1,1+2n}$  gauge group. 

\begin{table}[H]\caption{The 3d $\mathcal{N}=2$ $U(2)_{1,1+2n}$ gauge theory with $2(\, {\tiny \protect\yng(1)}_{\, 1}+ {\tiny \overline{\protect\yng(1)}}_{\,-1})$} 
\begin{center}
\scalebox{1}{
  \begin{tabular}{|c||c|c|c|c|c|c| } \hline
  &$U(2)_{1,1+2n}$&$SU(2)$&$SU(2)$&$U(1)_A$&$U(1)_T$&$U(1)_R$ \\ \hline
$Q$& ${\tiny \yng(1)}_{\, 1}$&${\tiny \yng(1)}$&1&$1$&0&$r$ \\  
$\tilde{Q}$& ${\tiny \overline{\yng(1)}}_{\, -1}$&1&${\tiny \yng(1)}$&1&0&$r$  \\ \hline
$M:= Q\tilde{Q}$&$\mathbf{1}$&${\tiny \yng(1)}$&${\tiny \yng(1)}$&2&0&$2r$  \\ \hline
  \end{tabular}}
  \end{center}\label{GGKU(2)F=2ele}
\end{table}

The magnetic description becomes a 3d $\mathcal{N}=2$ $\stackrel{+1~~~~~~}{\protect\wick{1}{<1 U(1)_{n+1} \times >1 U(1)_{0} }}$ gauge theory with two flavors and a gauge-singlet meson $M$. There is a level-1 mixed CS term between the two $U(1)$ gauge groups. The theory has a tree-level superpotential $W_{mag}=Mq \tilde{q}$ whose $F$-flatness condition lifts the dual meson $q \tilde{q}$. The meson singlet $M$ is identified with the composite $Q\tilde{Q}$ as usual. Since the second $U(1)$ gauge group has no CS term, one might consider that there is a Coulomb moduli space associated with the second $U(1)$ factor for any $n$. However, this is not the case because the mixed CS term gives the bare monopole operators non-zero charges under the first $U(1)$ gauge group and we cannot dress it. For $n=-1$, the first $U(1)$ gauge group also has zero CS level and the corresponding Coulomb branch can be gauge-invariant by using $q$ or $\tilde{q}$. Notice that the dual meson $q \tilde{q}$ is removed from the chiral ring elements but we can turn on $q$ or $\tilde{q}$, individually. This picture is nicely consistent with the analysis on the electric side.

\begin{table}[H]\caption{The magnetic $\stackrel{+1~~~~~~}{\protect\wick{1}{<1 U(1)_{n+1} \times >1 U(1)_{0} }}$ gauge theory dual to Table \ref{GGKU(2)F=2ele}} 
\begin{center}
\scalebox{1}{
  \begin{tabular}{|c||c|c|c|c|c|c| } \hline
  &$\stackrel{+1~~~~~~}{\protect\wick{1}{<1 U(1)_{n+1} \times >1 U(1)_{0} }}$&$SU(2)$&$SU(2)$&$U(1)_A$&$U(1)_T$&$U(1)_R$ \\ \hline
$q$& $(0,1)$&${\tiny \overline{\yng(1)}}$&1&$-1$&0&$1-r$ \\  
$\tilde{q}$& $(0,-1)$&1&${\tiny \overline{\yng(1)}}$&$-1$&0&$1-r$  \\ 
$M$&$\mathbf{1}$&${\tiny \yng(1)}$&${\tiny \yng(1)}$&2&0&$2r$  \\ \hline
  \end{tabular}}
  \end{center}\label{GGKU(2)F=2mag}
\end{table}

\subsubsection*{The case with $n=2$}
We here compute the superconformal indices by using the electric and magnetic descriptions for the $n=2$ case. We observed a nice agreement between the electric and magnetic indices. The indices for $n=2$ are given by  
\begin{align}
\left.  I^{U(2)_{1,1+2n}}_{(F,\bar{F})=(2,2)}  \right|_{n=2}&=1+4 t^2 \sqrt{x}+10 t^4 x+20 t^6 x^{3/2}+\left(35 t^8-8\right) x^2+\left(56 t^{10}-24 t^2\right) x^{5/2} \nonumber \\
&\qquad +\left(84 t^{12}-48 t^4-\frac{1}{t^4}\right) x^3+\left(120 t^{14}-80 t^6+\frac{8}{t^2}\right) x^{7/2} \nonumber \\
&\qquad +\left(165 t^{16}-120 t^8+28\right) x^4 +\left(220 t^{18}-168 t^{10}+32 t^2\right) x^{9/2} \nonumber \\
&\qquad  \qquad +\left(286 t^{20}-224 t^{12}+20 t^4-\frac{8}{t^4}\right) x^5+\cdots,
\end{align}
where $t$ is a fugacity parameter for the axial $U(1)_A$ symmetry. Although the r-charge of $Q$ and $\tilde{Q}$ is here set to be $r=\frac{1}{4}$ for simplicity, one can easily check the agreement of the indices for different values of $r$. The meson singlet $M:=Q\tilde{Q}$ is represented by $4 t^2 \sqrt{x}$ in the above expansion. The higher-order terms are the symmetric products of $M$ and the fermionic contributions. For example, the negative contribution $-8 x^2$ comes from the fermion composites $Q \psi_Q+\tilde{Q}\psi_{\tilde{Q}}$, where $\psi$ denotes the fermion partner of the squark. The leading monopole operator is represented as a fermion state $-\frac{1}{t^4}x^3$. This corresponds to $\mathbf{1}_{0,-2}\ket{1,-1} \sim W_\alpha \ket{1,0}$, where the state with a GNO charge $(1,-1)$ is denoted by $\ket{1,-1}$. $\mathbf{1}_{0,-2}$ denotes one component of the gaugino field $W_\alpha$. The insertion of the state $\ket{1,-1}$ breaks the gauge group as $U(2)_{1,5} \cong U(1)_{5} \times SU(2)_{1} \rightarrow U(1)_5 \times U(1)_{2}$. The $U(1)_2$ charge of the state $\ket{1,-1}$ is canceled by the gaugino field $W_\alpha$.

\subsubsection*{The case with $n=1$}
For $n=1$, the electric gauge group is $U(2)_{1,3}$. The Coulomb moduli space is not allowed since there is no flat direction where these CS levels are canceled. The theory only has the Higgs moduli space parametrized by $M:=Q\tilde{Q}$. Therefore, the superconformal index for $n=1$ is similar to the previous one. The electric and magnetic superconformal indices for $n=1$ are computed as
\begin{align}
\left.  I^{U(2)_{1,1+2n}}_{(F,\bar{F})=(2,2)}  \right|_{n=1}&=1+4 t^2 \sqrt{x}+10 t^4 x+20 t^6 x^{3/2}+\left(35 t^8-8\right) x^2+\left(56 t^{10}-24 t^2\right) x^{5/2} \nonumber \\
&\qquad+\left(84 t^{12}-48 t^4-\frac{1}{t^4}\right) x^3+\left(120 t^{14}-80 t^6+\frac{8}{t^2}\right) x^{7/2}+4 t^3 x^{15/4} \nonumber \\
&\qquad+\left(165 t^{16}-120 t^8+28\right) x^4+12 t^5 x^{17/4}+\left(220 t^{18}-168 t^{10}+32 t^2\right) x^{9/2} \nonumber \\
&\qquad \qquad +24 t^7 x^{19/4}+\left(286 t^{20}-224 t^{12}+20 t^4-\frac{8}{t^4}\right) x^5+\cdots
\end{align}
where the fugacity parameter of the axial $U(1)_A$ symmetry is denoted by $t$. The r-charge assignment is the same as the previous one. We observed this agreement up to $O(x^5)$. As a consistency check between the electric and magnetic indices, let's focus on the monopole state which is represented by $4 t^3 x^{15/4}$. On the electric side, the insertion of the bare monopole, which is denoted by $X_{\pm}$, induces the gauge symmetry breaking 
\begin{align}
U(2)_{1,3} &\rightarrow  \stackrel{+1~~~~~~~~}{\protect\wick{1}{<1 U(1)_{2} \times >1 U(1)_{2} }}\\
  {\tiny \yng(1)}_{\,1}  & \rightarrow  (+1, 0)+ (0, +1)\\ 
{\tiny \overline{\yng(1)}}_{\,-1}  & \rightarrow  (-1, 0) +(0,-1) \\
\mathbf{adj.}_{\, 0}& \rightarrow  (0,0)+(0,0)+(+1,-1)+(-1,+1),
\end{align}
where $\mathbf{adj.}_{\, 0}$ is a gaugino field $W_\alpha$.
Since the bare operator $X_+$ has the charge $(-2,-1)$ under the $U(1)_2 \times U(1)_2$ unbroken subgroup, the leading gauge-invariant state seems to be
\begin{align}
X_{+} (+1,0)^2 (0,1) \sim X_{+} Q^2 Q.
\end{align}
However, we should notice that the two states, $(+1,0) \sim Q$ and $(+1,-1)(0,1) \sim Q W_\alpha$, have the same quantum numbers with opposite spin-statistics. Therefore, the following state
\begin{align}
X_{+} (+1,0) (+1,-1) (0,+1) (0,+1)  \sim X_{+} Q (W_\alpha Q ) Q
\end{align}
precisely cancels out the (would-be) leading (bosonic) monopole state. The genuine leading states can be observed as
\begin{align}
X_{+} (+1,0)^2 (0,+1)^2 (0,-1) &\sim X_{+}Q^2 Q^2 \tilde{Q} \\
X_{+} (+1,0) (+1,-1) (0,+1)^3 (0,-1) &\sim X_+ QW_\alpha Q^3 \tilde{Q}
\end{align}
The composite state in the first line is a boson and has $18$ components under the non-abelian flavor symmetry while the second one is fermionic and has $16$ components. Therefore, the two boson states remain. We can give a similar argument for $X_-$ as well. This explains the leading contribution of the dressed monopole $4 t^3 x^{15/4}$.

On the magnetic side, these monopole states can be more easily detected: The magnetic gauge group becomes $U(1)_2 \times U(1)_0$ for $n=1$. There is also a level-1 mixed CS term. For these two $U(1)$ gauge groups, there are two sets of monopole operators which are denoted by $\tilde{\mathcal{U}}_{\pm}$ and $\tilde{X}_{\pm}$. Due to the (mixed) CS terms, these bare operators are charged under the topological $U(1)_2$ symmetry. The possible gauge-invariant states are determined as
\begin{align}
\tilde{\mathcal{U}}_+ \tilde{X}^2_- q,~~~~~\tilde{\mathcal{U}}_- \tilde{X}_+^2 \tilde{q},  
\end{align}
which are represented by $4 t^3 x^{15/4}$ in the superconformal indices. This observation confirms the validity of our duality proposal. 

\subsubsection*{The case with $n=0$}
For $n=0$, which is a conventional Giveon-Kutasov duality, the electric gauge group becomes $U(2)_{1,1}$. The leading monopole operator $X_{\pm}$ that appears in the superconformal index is associated with the gauge symmetry breaking $U(2)_{1,1} \rightarrow U(1)_1 \times U(1)_1$. Due to the CS term, we need to define gauge-invariant dressed states. The (would-be) leading state becomes
\begin{align}
X_{+} (1,0) \sim X_+ Q,~~~~X_{+} (1,-1) (0,1) \sim X_{+}W_{\alpha} Q,
\end{align}
which have the same quantum numbers except for their spin-statistics. As a result, these two states are canceled with each other. The non-vanishing states appear as
\begin{align}
X_{+} (1,0) (0,1) (0,-1) &\sim X_+Q^2\tilde{Q}\\
X_{+} (1,-1)(0,1)^2(0,-1) &\sim X_+ W_\alpha Q^2 \tilde{Q}
\end{align}
From the flavor structure of these composite states, we find that they are represented as $(-8+6) t x^{2+r}$ in the superconformal indices. For the other monopole $X_-$, the same argument is applicable. As a result, the leading dressed monopoles appear as $-4 tx^{2+r}$.

On the magnetic side, the gauge group is $U(1)_1 \times U(1)_0$. We denote the monopole for the $U(1)_1$ subgroup by $\tilde{\mathcal{U}}_{\pm}$ and the monopole for the $U(1)_0$ subgroup by $\tilde{X}_{\pm}$. The dressed gauge-invariant states are defined as
\begin{align}
\tilde{\mathcal{U}}_+ \tilde{X}_- q,~~~~\tilde{\mathcal{U}}_- \tilde{X}_+ \tilde{q},
\end{align}
which are represented as $-4 tx^{2+r}$ in the superconformal indices. As a consistency check of the duality for $n=0$, we compute the electric and magnetic indices. The result is expanded as 
\begin{align}
\left.  I^{U(2)_{1,1+2n}}_{(F,\bar{F})=(2,2)}  \right|_{n=0}&=1+4 t^2 \sqrt{x}+10 t^4 x+20 t^6 x^{3/2}+\left(35 t^8-8\right) x^2-4 t x^{9/4}  \nonumber \\
&\qquad+\left(56 t^{10}-24 t^2\right) x^{5/2}-12 t^3 x^{11/4} +\left(84 t^{12}-48 t^4-\frac{1}{t^4}\right) x^3-24 t^5 x^{13/4}   \nonumber \\
&\qquad+\left(120 t^{14}-80 t^6+\frac{8}{t^2}\right) x^{7/2}+\left(\frac{16}{t}-40 t^7\right) x^{15/4}   \nonumber \\
&\qquad+\left(165 t^{16}-120 t^8+28\right) x^4+\left(44 t-60 t^9\right) x^{17/4}  \nonumber \\
&\qquad+\left(220 t^{18}-168 t^{10}+32 t^2\right) x^{9/2}  \nonumber \\
&\qquad \qquad +\left(68 t^3-84 t^{11}\right) x^{19/4}+\left(286 t^{20}-224 t^{12}+20 t^4-\frac{8}{t^4}\right) x^5+\cdots,
\end{align}
where $t$ is a fugacity parameter for the $U(1)_A$ symmetry. The r-charge is set to be $r=\frac{1}{4}$ for simplicity. We observed this agreement up to $O(x^5)$. The second term $4 t^2 \sqrt{x}$ corresponds to the meson $M:=Q \tilde{Q}$. The dressed monopoles are represented by $-4 tx^{2+r}|_{r=\frac{1}{4}}=-4 t x^{9/4} $. 

\subsubsection*{The case with $n=-1$}
For $n=-1$, the electric gauge group becomes $U(2)_{1,-1}$, which allows a Coulomb moduli space: Along the Coulomb branch, whose coordinate is denoted by $X_{\pm}$, the gauge group is spontaneously broken to $U(1)_0 \times U(1)_0$. Since the CS levels are canceled, this flat direction becomes exactly massless. Due to the mixed CS terms between the two $U(1)_0$ subgroups, the bare operator $X_{\pm}$ obtains a non-zero $U(1)_0$ charge. Then, we need to define the dressed monopole operators
\begin{align}
X_{+d} &:=X_+ (0,-1)\sim X_+ \tilde{Q} \\
X_{-d} &:=X_-(0,1) \sim X_- Q.
\end{align}

On the magnetic side, the gauge group becomes $U(1)_0 \times U(1)_0$. Since the two $U(1)$ vector multiplets have no CS term (except for the mixed CS term), there are two Coulomb flat directions: Let us denote the monopole operator for the first $U(1)_0$ subgroup by $\tilde{\mathcal{U}}_{\pm}$ and the latter one by $\tilde{X}_{\pm}$. Due to the mixed CS term, these bare monopoles are not gauge-invariant. Therefore, we need to consider the following composites
\begin{align}
\tilde{\mathcal{U}}_{-} \tilde{q},~~~~~\tilde{\mathcal{U}}_{+} q.
\end{align}
These are identified with $X_{\pm d}$. Notice that the bare operator $\tilde{X}_{\pm}$ is charged under the first $U(1)_0$ symmetry and cannot be made gauge-invariant by matter multiplets. Although the composite $\tilde{X}_{+} \tilde{X}_-$ is gauge-invariant, we cannot simultaneously turn on these operators since $\tilde{X}_{+}$ and $\tilde{X}_-$ parametrize positive and negative eigenvalues of the adjoint scalar, respectively. On the magnetic side, we can more easily study the dressed operators: Since the matter fields are not charged under the (gauged) topological $U(1)_{top}$ symmetry and the $U(1)_{top}$ dynamics only includes the BF coupling, we can integrate over the $U(1)_{top}$ vector multiplet. The resulting theory is just a non-gauge theory with the three gauge-singlets $q$, $\tilde{q}$ and $M$. In this picture, the electric dressed monopoles $X_{\pm d}$ are simply mapped to $q$ and $\tilde{q}$.

As a test of the above analysis, we compute the electric and magnetic superconformal indices. We observed a nice agreement and the result is expanded as 
\begin{align}
\left.  I^{U(2)_{1,1+2n}}_{(F,\bar{F})=(2,2)}  \right|_{n=-1}&=1+4 t^2 \sqrt{x}+\frac{4 x^{3/4}}{t}+10 t^4 x+12 t x^{5/4}+\left(20 t^6+\frac{6}{t^2}\right) x^{3/2}+24 t^3 x^{7/4}  \nonumber \\
&\qquad  +\left(35 t^8+8\right) x^2  +\left(40 t^5+\frac{4}{t^3}\right) x^{9/4}+\left(56 t^{10}+6 t^2\right) x^{5/2}   \nonumber \\
&\qquad +\left(60 t^7-\frac{4}{t}\right) x^{11/4}+\left(84 t^{12}+\frac{1}{t^4}\right) x^3  +\left(84 t^9-4 t\right) x^{13/4} \nonumber \\
&\qquad +\left(120 t^{14}-10 t^6\right) x^{7/2}+\left(112 t^{11}+4 t^3\right) x^{15/4} +\left(165 t^{16}-24 t^8+22\right) x^4   \nonumber \\
&\qquad +\left(144 t^{13}+20 t^5+\frac{4}{t^3}\right) x^{17/4}+\left(220 t^{18}-42 t^{10}+32 t^2\right) x^{9/2}  \nonumber \\
&\qquad \quad +\left(180 t^{15}+44 t^7-\frac{12}{t}\right) x^{19/4}+\left(286 t^{20}-64 t^{12}+30 t^4-\frac{8}{t^4}\right) x^5+\cdots \\
&=\left.  \left( \frac{(t^{-2} x^{2-2r}; x^2 )_\infty}{(t^2 x^{2r};x^2)_\infty} \right)^4  \left( \frac{(t^{1} x^{2-(1-r)}; x^2 )_\infty}{(t^{-1} x^{1-r};x^2)_\infty} \right)^2  \left( \frac{(t^{1} x^{2-(1-r)}; x^2 )_\infty}{(t^{-1} x^{1-r};x^2)_\infty} \right)^2 \right|_{r=\frac{1}{4}},
\end{align}
where $(a;q)_{\infty} :=\prod_{k=0}^{\infty} (1-aq^k)$ is a $q$-Pochhammer symbol. The parameter $t$ denotes the fugacity for the axial $U(1)_A$ symmetry. The r-charge is set to be $r=\frac{1}{4}$ for simplicity. In the last line above, the indices are reorganized into the indices of the three gauge-singlets $M$, $q$ and $\tilde{q}$, which confirms that the theory exhibits s-confinement.  

\subsubsection*{The case with $n=-2$}
For $n=-2$, the electric gauge group becomes $U(2)_{1,-3}$. The theory has no Coulomb moduli space. As a test of the duality, we examine the leading monopole operator which appears in the superconformal indices. The insertion of the monopole operator $X_{\pm}$ induces the gauge symmetry breaking $U(2)_{1,-3} \rightarrow \stackrel{-2~~~~~~~~}{\protect\wick{1}{<1 U(1)_{-1} \times >1 U(1)_{-1} }}$. Since the CS terms are not vanishing, this direction of the Coulomb branch is eliminated from the quantum moduli space but contributes to the SCI. Due to the (mixed) CS terms, the bare operators $X_{\pm}$ are charged under the $U(1) \times U(1)$ unbroken subgroup. The gauge-invariant states that appear in the SCI is determined as
\begin{align}
X_{+d}:= X_+ (-1,0) (0,-1)^2 \sim X_+ \tilde{Q} \tilde{Q}^2 \\
X_{-d}:= X_- (+1,0) (0,+1)^2 \sim X_+ Q Q^2,
\end{align}
where the components $(-1,0)$ and $(+1,0)$ behave as fermions on the monopole background. As a result, these dressed monopoles are fermionic. 

On the magnetic side, the gauge group becomes $\stackrel{+1~~~~\,~}{\protect\wick{1}{<1 U(1)_{-1} \times >1 U(1)_{0} }}$. For the $U(1)_{-1}$ subgroup, the monopole operator is denoted by $\tilde{\mathcal{U}}_{\pm}$ while the second $U(1)_0$ monopole is represented by $\tilde{X}_{\pm}$. Although the CS level of the second gauge group is vanishing, there is no associated Coulomb branch. This is because the bare operator $\tilde{X}_{\pm}$ is charged under the $U(1)_{-1}$ symmetry due to the mixed CS term. The leading monopole operator is defined as
\begin{align}
\tilde{\mathcal{U}}_+ \tilde{X}_+ q  \\
\tilde{\mathcal{U}}_- \tilde{X}_-  \tilde{q}. 
\end{align}
On the monopole background $\tilde{X}_{\pm}$, the spins of $q$ and $\tilde{q}$ are transmuted to fermion-statistics \cite{Polyakov:1988md, Dimofte:2011py, Aharony:2015pla}. These dressed states are identified with $X_{\pm d}$ under the duality. 

As a consistency check of the above analysis, let us compute the superconformal indices. On both the electric and magnetic side, the indices are expanded as
\begin{align}
\left.  I^{U(2)_{1,1+2n}}_{(F,\bar{F})=(2,2)}  \right|_{n=-2}&=1+4 t^2 \sqrt{x}+10 t^4 x+20 t^6 x^{3/2}+\left(35 t^8-8\right) x^2 -4 t x^{9/4}  \nonumber \\
&\qquad +\left(56 t^{10}-24 t^2\right) x^{5/2}   -12 t^3 x^{11/4} +\left(84 t^{12}-48 t^4-\frac{1}{t^4}\right) x^3  \nonumber \\
&\qquad-24 t^5 x^{13/4}+\left(120 t^{14}-80 t^6+\frac{8}{t^2}\right) x^{7/2}  +\left(\frac{16}{t}-40 t^7\right) x^{15/4} \nonumber \\
&\qquad+\left(165 t^{16}-120 t^8+28\right) x^4+\left(44 t-60 t^9\right) x^{17/4}  \nonumber \\
&\qquad+\left(220 t^{18}-168 t^{10}+32 t^2\right) x^{9/2}+\left(68 t^3-84 t^{11}\right) x^{19/4}  \nonumber \\
&\qquad \qquad +\left(286 t^{20}-224 t^{12}+20 t^4-\frac{8}{t^4}\right) x^5+\cdots,
\end{align}
where the fugacity parameter $t$ is introduced only for the axial $U(1)_A$ symmetry and the r-charge is set to be $r=\frac{1}{4}$ for simplicity. The second term $4 t^2 \sqrt{x}$ represents a meson $M:=Q \tilde{Q}$. The sixth term $-4 t x^{9/4} $ is the dressed monopole $X_{\pm d}$. The higher-order terms can be regarded as symmetric products of these fields, the fermion contributions and more complicated monopole operators.  

\subsubsection*{The case with $n=-3$}
Finally, we study the $n=-3$ case. Although there is no Coulomb moduli space, the theory has the monopole operators which appear in the SCI. As a test of the duality, we consider the matching of the dressed monopoles under the duality. We find that the superconformal indices on both the electric and magnetic sides are computed as
\begin{align}
\left.  I^{U(2)_{1,1+2n}}_{(F,\bar{F})=(2,2)}  \right|_{n=-3}&=1+4 t^2 \sqrt{x}+10 t^4 x+20 t^6 x^{3/2}+\left(35 t^8-8\right) x^2+\left(56 t^{10}-24 t^2\right) x^{5/2}  \nonumber \\
&\qquad+\left(84 t^{12}-48 t^4-\frac{1}{t^4}\right) x^3+\left(120 t^{14}-80 t^6+\frac{8}{t^2}\right) x^{7/2}+4 t^3 x^{15/4}  \nonumber \\
&\qquad+\left(165 t^{16}-120 t^8+28\right) x^4+12 t^5 x^{17/4}+\left(220 t^{18}-168 t^{10}+32 t^2\right) x^{9/2}  \nonumber \\
&\qquad \qquad +24 t^7 x^{19/4}+\left(286 t^{20}-224 t^{12}+20 t^4-\frac{8}{t^4}\right) x^5+\cdots,
\end{align}
where the fugacity for the $U(1)_A$ symmetry is denoted by $t$ and the r-charge is set to be $r=\frac{1}{4}$ for simplicity. The meson composite $M:=Q \tilde{Q}$ is represented as $4 t^2 \sqrt{x}$ as in the previous examples. We here give an operator interpretation of the ninth term $4 t^3 x^{15/4}$ in the above expansion, which is regarded as a leading monopole state: On the electric side, the insertion of the bare monopole denoted by $X_{\pm}$ induces the gauge symmetry breaking
\begin{align}
U(2)_{1,-5} & \rightarrow \stackrel{-3~~~~~~~~}{\protect\wick{1}{<1 U(1)_{-2} \times >1 U(1)_{-2} }} \\
 {\tiny \yng(1)}_{\,1}  & \rightarrow  (+1, 0)+ (0, +1)\\ 
{\tiny \overline{\yng(1)}}_{\,-1}  & \rightarrow  (-1, 0) +(0,-1) ,
\end{align}
where the monopole is associated with the first $U(1)_{-2}$ subgroup. 
Due to the (mixed) CS terms, the bare monopole $X_{\pm}$ is charged under the $U(1)_{-2} \times U(1)_{-2}$ subgroup. As a result, the gauge-invariant states are determined as
\begin{align}
X_{+d} &:= X_{+} (-1,0)^2 (0,-1)^3 \sim X_+ \tilde{Q}^2 \tilde{Q}^3 \\
X_{- d} &:=X_- (+1,0)^2 (0,+1)^3  \sim X_- Q^2 Q^3,
\end{align}
whose quantum numbers correctly explain the ninth term $4 t^3 x^{15/4}$. Note that the components $(-1,0)$ and $(+1,0)$ behave as fermions on the monopole $X_{\pm}$ background \cite{Polyakov:1988md, Dimofte:2011py, Aharony:2015pla}. As a whole, the above combinations contribute to the indices as bosons.

On the magnetic side, the gauge group becomes $\stackrel{+1~~~~~~}{\protect\wick{1}{<1 U(1)_{-2} \times >1 U(1)_{0} }}$ for $n=-3$. We denote the monopole associated with the first $U(1)_{-2}$ gauge group by $\tilde{\mathcal{U}}_{\pm}$ while the monopole associated with the second $U(1)_0$ gauge group is denoted by $\tilde{X}_{\pm}$. Although the second $U(1)_0$ has zero CS level, there is no Coulomb moduli space because the mixed CS term gives a non-zero $U(1)_{-2}$ charge to $\tilde{X}_{\pm}$ and its charge cannot be canceled by the massless matter fields. These monopole operators are not gauge-invariant due to the various CS terms but we can define the dressed monopoles as
\begin{align}
\tilde{\mathcal{U}}_+ \tilde{X}_+^2 q~~~~\mbox{and}~~~~\tilde{\mathcal{U}}_- \tilde{X}_-^2 \tilde{q},
\end{align}
which are gauge-invariant and identified with $X_{-d}$ and $X_{+d}$, respectively.

\subsection{$U(1)_{0}$ with a single flavor}
Next, we examine the abelian example. Namely, we will take $N=1$. This case is very special because the electric gauge group becomes $U(1)_{\tilde{k}=k+n}$ and its CS level is parametrized by a single integer $\tilde{k}$. The dual gauge group becomes $U(1)_{n+1} \times U(F+|\tilde{k}-n|-N)_{n-\tilde{k}, F-N}$, which is labeled by $\tilde{k}$ and $n$. This means that there are an infinite number of magnetic descriptions with a free parameter $n$ for one electric theory. This is an example of the duality enhancement.

We here consider the 3d $\mathcal{N}=2$ $U(1)_0$ gauge theory with one flavor, which is known to be dual to the 3d $\mathcal{N}=2$ XYZ model \cite{Aharony:1997bx, deBoer:1997kr}. Therefore, the electric and magnetic theories exhibit s-confinement. The theory can be obtained by taking $N=1$ and $n=-k$ in our duality. The Higgs branch is described by a meson field $M:=Q \tilde{Q}$. The Coulomb branch is split into two regions and parametrized by $V_{\pm}$ \cite{Aharony:1997bx, deBoer:1997kr}. Table \ref{XYZele} summarizes the quantum numbers of these moduli coordinates. The low-energy effective description is dual to a non-gauge theory with three (gauge-singlet) chiral multiplets and a cubic superpotential
\begin{align}
W_{eff}= MV_+V_-.
\end{align}

\begin{table}[H]\caption{The 3d $\mathcal{N}=2$ $U(1)_{0}$ gauge theory with one flavor} 
\begin{center}
\scalebox{1}{
  \begin{tabular}{|c||c|c|c|c| } \hline
  &$U(1)_{0}$&$U(1)_A$&$U(1)_T$&$U(1)_R$ \\ \hline
$Q$& $+1$&$1$&0&$r$ \\  
$\tilde{Q}$& $-1$&1&0&$r$  \\ \hline
$M:= Q\tilde{Q}$&$0$&2&0&$2r$  \\ \hline
$V_{\pm}$&0&$-1$&$\pm 1$&$1-r$  \\ \hline
  \end{tabular}}
  \end{center}\label{XYZele}
\end{table}

The magnetic side becomes a 3d $\mathcal{N}=2$ $\stackrel{+1~~~~~~~~~}{\protect\wick{1}{<1 U(1)_{-k+1} \times >1 U(k)_{-k,0} }}$ gauge theory with a dual flavor and a meson singlet $M$. The theory has a tree-level superpotential $W_{mag} =Mq \tilde{q}$. The dual descriptions are available for any integer $k$ and exhibit duality enhancement. The dual meson $q\tilde{q}$ is removed from the moduli space by the $F$-flatness condition of the meson $M$. Although the abelian CS level for the $U(k)_{-k,0}$ gauge group is zero, the corresponding $U(1)$ direction cannot be a part of the moduli space. This is because the mixed CS term makes the bare monopole operator gauge non-invariant under the $U(1)_{-k+1}$ subgroup and this charge cannot be canceled by the matter multiplets. Therefore, we have to consider a more non-trivial Coulomb flat direction. 
For the magnetic Coulomb moduli space, we will consider the following gauge symmetry breaking
\begin{align}
U(1)_{-k+1} \times  U(k)_{-k,0} & \rightarrow  U(1)_{-k+1} \times U(k-1)_{-k,-1} \times U(1)_{-k+1} \\
(0,{\tiny \yng(1)}_{\,1} ) & \rightarrow (0,{\tiny \yng(1)}_{\,1},0 )+(0,\mathbf{1}_0,+1) \\
(0,{\tiny \overline{\yng(1)}}_{\,-1})  & \rightarrow (0,{\tiny \overline{\yng(1)}}_{\,-1},0) +(0, \mathbf{1}_0,-1)
\end{align}
where the breaking is induced by the monopole operators $\tilde{X}_{\pm}$ which corresponds to the $U(1) \subset U(k-1)$ generator. The level-$1$ mixed CS terms are introduced for all the combinations of the $U(1)$ subgroups. We denote the monopole operators for the topological $U(1)_{-k+1}$ subgroup by $\tilde{\mathcal{U}}_{\pm}$. Since the CS terms are introduced for all the $U(1)$ gauge groups, it seems that the above Coulomb branch becomes massive. However, by simultaneously turning on both the $U(1)_{-k+1}$ and $U(1) \subset U(k-1)_{-k-1}$ Coulomb branches, the following monopole operators can be a part of the Coulomb moduli space
\begin{align}
\tilde{\mathcal{U}}_+ \tilde{X}_+^{k-1},~~~~~~\tilde{\mathcal{U}}_- \tilde{X}_-^{k-1}.
\end{align}
Due to the mixed CS terms, these bare monopoles are charged under the $U(1)_{-k+1}$ subgroup \cite{Intriligator:2013lca}. Therefore, we need to consider the dressed Coulomb branch coordinates
\begin{align}
V_+ &:= \tilde{\mathcal{U}}_+ \tilde{X}_+^{k-1} (0,\mathbf{1}_0,+1)^{k} \sim \tilde{\mathcal{U}}_+ \tilde{X}_+^{k-1} q^k  \\
V_- &:= \tilde{\mathcal{U}}_- \tilde{X}_-^{k-1} (0, \mathbf{1}_0,-1)^k  \sim \tilde{\mathcal{U}}_- \tilde{X}_-^{k-1} \tilde{q}^k,
\end{align}
which are identified with the electric monopole operators $V_{\pm}$. The low-energy dynamics is described by three gauge-invariants $M$ and $V_{\pm}$ with a cubic superpotential $W_{eff}=MV_+V_-$, which is consistent with all the symmetries.

\begin{table}[H]\caption{The magnetic $\stackrel{+1~~~~~~~~~}{\protect\wick{1}{<1 U(1)_{-k+1} \times >1 U(k)_{-k,0} }}$ gauge theory dual to Table \ref{XYZele}} 
\begin{center}
\scalebox{1}{
  \begin{tabular}{|c||c|c|c|c| } \hline
  &$\stackrel{+1~~~~~~~~~}{\protect\wick{1}{<1 U(1)_{-k+1} \times >1 U(k)_{-k,0} }}$&$U(1)_A$&$U(1)_T$&$U(1)_R$ \\ \hline
$q$& $(0,{\tiny \yng(1)}_{\,1} )$&$-1$&0&$1-r$ \\  
$\tilde{q}$& $(0,{\tiny \overline{\yng(1)}}_{\,-1})$&$-1$&0&$1-r$  \\
$M$&$0$&2&0&$2r$  \\ \hline
$V_+ \sim \tilde{\mathcal{U}}_+ \tilde{X}_+^{k-1} q^k$&0&$-1$&$+1$&$1-r$  \\
$V_- \sim \tilde{\mathcal{U}}_- \tilde{X}_-^{k-1} \tilde{q}^k$&0&$-1$&$-1$&$1-r$  \\ \hline
  \end{tabular}}
  \end{center}\label{XYZmag}
\end{table}

For $k=1$, we verified the agreement of the electric and magnetic superconformal indices. The result is expanded as 
\begin{align}
I_{(F,\bar{F})=(1,1)}^{U(1)_{0}} &=1+t^2 \sqrt{x}+\frac{2 x^{3/4}}{t}+t^4 x+\left(t^6+\frac{2}{t^2}\right) x^{3/2}+\left(t^8-2\right) x^2+\frac{2 x^{9/4}}{t^3}+t^{10} x^{5/2} \nonumber \\
&\qquad +\left(t^{12}+\frac{2}{t^4}\right) x^3+2 t x^{13/4}+\left(t^{14}+\frac{1}{t^2}\right) x^{7/2}+\frac{2 x^{15/4}}{t^5}+\left(t^{16}-3\right) x^4+\cdots  \nonumber \\
&=\left.   \frac{(t^{-2} x^{2-2r}; x^2 )_\infty}{(t^2 x^{2r};x^2)_\infty}  \cdot  \frac{(t^{1} x^{2-(1-r)}; x^2 )_\infty}{(t^{-1} x^{1-r};x^2)_\infty}  \cdot  \frac{(t^{1} x^{2-(1-r)}; x^2 )_\infty}{(t^{-1} x^{1-r};x^2)_\infty}  \right|_{r=\frac{1}{4}},
\end{align}
where the r-charge is set to $r=\frac{1}{4}$ for simplicity and the parameter $t$ denotes the fugacity for the axial $U(1)_A$ symmetry. The meson $M:=Q\tilde{Q}$ is represented by $t^2 \sqrt{x}$ and the two monopole operators $V_{\pm}$ are denoted as $\frac{2 x^{3/4}}{t}$. In the last line, the indices are combined into the indices of the three gauge-singlet chiral superfields $M$, $V_+$ and $V_-$. This confirms the validity of our analysis.

\subsection{$U(1)_{k}$ with a single flavor}
The final example is a 3d $\mathcal{N}=2$ $U(1)_k$ gauge theory with one flavor. The electric and magnetic theories can be obtained by taking $N=k=1$ and relabeling $n\rightarrow k-1$ in our duality proposal. One can easily generalize the analysis in this subsection by adding many flavors. We here only consider the $F=1$ case for simplicity. The Higgs branch is described by the meson composite $M:=Q \tilde{Q}$. As studied in \cite{Benini:2011mf, Intriligator:2013lca}, the theory has no Coulomb branch since the bare CS level cannot be canceled for the vector-like matter content. As a consistency check of the duality, let us focus on the monopole operators which are not part of the moduli space but appear in the expansion of the superconformal indices. The bare monopole operators $V_{\pm}$ are charged under the $U(1)_k$ gauge symmetry due to the bare CS term. Hence, we need to define the dressed monopole states
\begin{align}
V_{+d}:=V_+Q^k,~~~~~~~V_{-d}:= V_- \tilde{Q}^k.
\end{align}
Table \ref{U(1)k1Fele} summarizes the quantum numbers of these operators. 
On the monopole $V_{\pm}$ background, the spins of the matter fields $Q$ and $\tilde{Q}$ are made fermionic \cite{Polyakov:1988md, Dimofte:2011py, Aharony:2015pla}. As a result, these dressed composites are bosonic states for even $k$ and are fermions for odd $k$. Since the theory has a singe CS term, we can also construct the regular Giveon-Kutasov dual theory whose gauge group is $U(k)_{-k,-k}$. In what follows, we only consider the generalized duality.

\begin{table}[H]\caption{The 3d $\mathcal{N}=2$ $U(1)_{k}$ gauge theory with one flavor} 
\begin{center}
\scalebox{1}{
  \begin{tabular}{|c||c|c|c|c| } \hline
  &$U(1)_{k}$&$U(1)_A$&$U(1)_T$&$U(1)_R$ \\ \hline
$Q$& $+1$&$1$&0&$r$ \\  
$\tilde{Q}$& $-1$&1&0&$r$  \\ \hline
$M:= Q\tilde{Q}$&$0$&2&0&$2r$  \\ \hline
$V_{\pm}$&$\mp k$&$-1$&$\pm 1$&$1-r$  \\ 
$V_{+d}:=V_+Q^k$&0&$k-1$&+1&$k+1+(k-1)r$  \\
$V_{-d}:= V_- \tilde{Q}^k$&0&$k-1$&$-1$&$k+1+(k-1)r$  \\ \hline
  \end{tabular}}
  \end{center}\label{U(1)k1Fele}
\end{table}

The magnetic description becomes a 3d $\mathcal{N}=2$ $\stackrel{+1~~~~~}{\protect\wick{1}{<1 U(1)_{k} \times >1 U(1)_{0} }}$ gauge theory with a dual flavor and a meson singlet $M:=Q \tilde{Q}$. The theory has a tree-level superpotential $W_{mag} =Mq \tilde{q}$. The first $U(1)_{k}$ gauge group is a topological $U(1)$ symmetry associated with the second $U(1)_0$ gauge group. The matter fields are not charged under the $U(1)_k$ topological symmetry. The quantum numbers of the elementary fields are summarized in Table \ref{U(1)k1Fmag}. We denote the first $U(1)_k$ monopoles by $\tilde{\mathcal{U}}_{\pm}$ and the monopoles for the second $U(1)_0$ group by $\tilde{X}_{\pm}$. Notice that the bare operators $\tilde{X}_{\pm}$ cannot be flat directions of the Coulomb moduli space although the corresponding CS level is zero. This is because the bare operators $\tilde{X}_{\pm}$ are charged under the $U(1)_k$ symmetry due to the mixed CS term. As a result, the theory does not allow any Coulomb branch even if one of the CS levels is vanishing. However, we can consider the dressed monopole operators which appear in the SCI: The leading dressed monopole states are defined as 
\begin{align}
\tilde{\mathcal{U}}_+ \tilde{X}_+^{k} q,~~~~~~~\tilde{\mathcal{U}}_- \tilde{X}_-^{k} \tilde{q}, \label{U(1)monopole_mag}
\end{align}
which contribute as fermions for odd $k$ and as bosons for even $k$. From the symmetry argument, these are identified with $V_{\pm d}$.

\begin{table}[H]\caption{The magnetic $\stackrel{+1~~~~~}{\protect\wick{1}{<1 U(1)_{k} \times >1 U(1)_{0} }}$ gauge theory dual to Table \ref{U(1)k1Fele}} 
\begin{center}
\scalebox{1}{
  \begin{tabular}{|c||c|c|c|c| } \hline
  &$\stackrel{+1~~~~~}{\protect\wick{1}{<1 U(1)_{k} \times >1 U(1)_{0} }}$&$U(1)_A$&$U(1)_T$&$U(1)_R$ \\ \hline
$q$& $(0,+1)$&$-1$&0&$1-r$ \\  
$\tilde{q}$& $(0,-1)$&$-1$&0&$1-r$  \\
$M$&$(0,0)$&2&0&$2r$  \\ \hline
$V_{+d} \sim \tilde{\mathcal{U}}_+ \tilde{X}_+^{k} q$&$(0,0)$&$k-1$&$+1$&$k+1+(k-1)r$  \\
$V_{-d} \sim \tilde{\mathcal{U}}_- \tilde{X}_-^{k} \tilde{q}$&$(0,0)$&$k-1$&$-1$&$k+1+(k-1)r$  \\ \hline
  \end{tabular}}
  \end{center}\label{U(1)k1Fmag}
\end{table}

For $k=2$, we will study the superconformal indices as a consistency check of our study. The indices of the electric and magnetic descriptions are expanded as 
\begin{align}
I_{(F,\bar{F})=(1,1)}^{U(1)_{k=2}} &=1+t^2 \sqrt{x}+t^4 x+t^6 x^{3/2}+\left(t^8-2\right) x^2+t^{10} x^{5/2}+t^{12} x^3+2 t x^{13/4} \nonumber \\
&\qquad +\left(t^{14}+\frac{1}{t^2}\right) x^{7/2}+\left(t^{16}-3\right) x^4+\left(t^{18}-2 t^2\right) x^{9/2}-\frac{2 x^{19/4}}{t}+t^{20} x^5+\cdots,
\end{align}
where the r-charges of $Q$ and $\tilde{Q}$ are chosen as $r=\frac{1}{4}$ and $t$ is a fugacity parameter for the $U(1)_A$ symmetry. The second term $t^2 \sqrt{x}$ is identified with the meson $M$. 
The leading monopole state appears as $2 t x^{13/4}$. This is regarded as the dressed monopoles $V_{\pm d}$ on the electric side. The same contribution can be recognized as \eqref{U(1)monopole_mag} on the magnetic side.  

\section{Summary and Discussion}
In this paper, we generalized the Giveon-Kutasov duality \cite{Giveon:2008zn} for the 3d $\mathcal{N}=2$ $U(N)$ gauge theory with $F$ fundamental flavors. The generalized gauge symmetry is $U(N)_{k,k+nN}$ where the Chern-Simon levels for the non-abelian and abelian subgroups are tuned to be $k$ and $k+n N$, respectively. For $N=1$, we pointed out that there are an infinite number of magnetic-dual theories.
The proposed duality is very similar to the non-supersymmetric bosonization duality proposed in \cite{Radicevic:2016wqn} and actually the duality discussed here is related to the non-supersymmetric one \cite{Radicevic:2016wqn} via a supersymmetry-breaking mass deformation. For several $n$, the duality reduces to the known dualities \cite{Choi:2018ohn, Aharony:2013dha, Aharony:2014uya}. For $N=1$ and $N=2$, we explicitly studied the duality for general $n$ and computed the superconformal indices as a validity test of the proposed duality. We discussed the matching of the leading monopole operator under the duality transformation. For almost all the choices of $n$, the monopole operator is not a part of the moduli space but appears in the SCI. For special values of $n$, this becomes a coordinate of the Coulomb moduli space.

We should note that the proposed duality is correct only for $k \neq 0$ and especially that the duality for $k =0 \, \cap \, n \neq 0$ is not known. For $k=n=0$, the $U(N)_{0,0}$ Seiberg-like duality was constructed by Aharony \cite{Aharony:1997gp}, which is now known as Aharony duality. The magnetic description of the Aharony duality includes additional gauge-singlets which become flat directions of the electric Coulomb branch. 
For the $SU(N)_{k=0}$ duality \cite{Aharony:2013dha}, which corresponds to the case with $k=0$ and $n= \infty$ in our duality, the magnetic description includes an electron-positron pair which is charged under the magnetic gauge group. Since the generalized duality connects these two dualities by changing the value of $n$, it is natural to think that the dual description for $k =0 \cap n \neq 0$ slightly differs from these two dualities. However, the fact that the additional fields are singlet in Aharony duality and charged in the $SU(N)_{k=0}$ duality makes it difficult to guess the correct duality for $k =0 \, \cap \, n \neq 0$. In our current understanding, we can only give the quantum structure of the Coulomb branch in the $U(N)_{0,nN}$ gauge theory: The theory allows the Coulomb moduli space where the gauge group is broken as $U(N) \cong U(1) \times SU(N) \rightarrow U(1) \times SU(N-2) \times U(1)_1 \times U(1)_2$. This flat direction is the same as the Coulomb branch of the vector-like $SU(N)_0$ SQCD theory \cite{Aharony:1997bx, deBoer:1997kr}. However, as opposed to the $SU(N)_0$ SQCD, the $U(N)_{0,nN}$ gauge theory does not allow the baryon operators. Due to this difference, we couldn't construct the duality for $k=0$ and $n\neq 0$ by mimicking the $SU(N)_{k=0}$ duality. We would like to go back to this problem in the foreseeable future.

As a future problem, it is important to study the derivation of this duality. Normally, almost all the 3d dualities are derived from other dualities: For example, the regular Giveon-Kutasov duality is connected to the Aharony duality via a real mass deformation \cite{Intriligator:2013lca, Khan:2013bba, Amariti:2013qea}. The 3d Seiberg-like dualities are related to the corresponding 4d dualities via dimensional reduction \cite{Gadde:2011ia, Imamura:2011uw, Dolan:2011rp, Niarchos:2012ah, Aharony:2013dha, Aharony:2013kma, Amariti:2015mva, Amariti:2016kat}. It would be valuable to derive the duality proposed here from other 3d/4d dualities. It is also important to consider the further generalization of the generalized Giveon-Kutasov duality with chiral matter content \cite{Benini:2011mf, Nii:2018bgf}. We hope that we can report on these problems elsewhere.

\section*{Acknowledgments}
Keita Nii would like to thank Shigeki Sugimoto and Seiji Terashima for valuable comments. Keita Nii is the Yukawa Research Fellow supported by Yukawa Memorial Foundation. This work was supported by JSPS KAKENHI Grant Number JP20K14466.

\bibliographystyle{ieeetr}
\bibliography{GGKduality}

\begin{thebibliography}{10}

\bibitem{Seiberg:1994bz}
N.~Seiberg, ``{Exact results on the space of vacua of four-dimensional SUSY
  gauge theories},'' {\em Phys. Rev.}, vol.~D49, pp.~6857--6863, 1994.

\bibitem{Seiberg:1994pq}
N.~Seiberg, ``{Electric - magnetic duality in supersymmetric nonAbelian gauge
  theories},'' {\em Nucl. Phys.}, vol.~B435, pp.~129--146, 1995.

\bibitem{Aharony:2013dha}
O.~Aharony, S.~S. Razamat, N.~Seiberg, and B.~Willett, ``{3d dualities from 4d
  dualities},'' {\em JHEP}, vol.~07, p.~149, 2013.

\bibitem{Aharony:2013kma}
O.~Aharony, S.~S. Razamat, N.~Seiberg, and B.~Willett, ``{3$d$ dualities from
  4$d$ dualities for orthogonal groups},'' {\em JHEP}, vol.~08, p.~099, 2013.

\bibitem{Intriligator:1995ax}
K.~A. Intriligator, R.~G. Leigh, and M.~J. Strassler, ``{New examples of
  duality in chiral and nonchiral supersymmetric gauge theories},'' {\em Nucl.
  Phys.}, vol.~B456, pp.~567--621, 1995.

\bibitem{Kutasov:1995ve}
D.~Kutasov, ``{A Comment on duality in N=1 supersymmetric nonAbelian gauge
  theories},'' {\em Phys. Lett.}, vol.~B351, pp.~230--234, 1995.

\bibitem{Kutasov:1995np}
D.~Kutasov and A.~Schwimmer, ``{On duality in supersymmetric Yang-Mills
  theory},'' {\em Phys. Lett.}, vol.~B354, pp.~315--321, 1995.

\bibitem{Intriligator:1995id}
K.~A. Intriligator and N.~Seiberg, ``{Duality, monopoles, dyons, confinement
  and oblique confinement in supersymmetric SO(N(c)) gauge theories},'' {\em
  Nucl. Phys.}, vol.~B444, pp.~125--160, 1995.

\bibitem{Intriligator:1995ne}
K.~A. Intriligator and P.~Pouliot, ``{Exact superpotentials, quantum vacua and
  duality in supersymmetric SP(N(c)) gauge theories},'' {\em Phys. Lett.},
  vol.~B353, pp.~471--476, 1995.

\bibitem{Giveon:2008zn}
A.~Giveon and D.~Kutasov, ``{Seiberg Duality in Chern-Simons Theory},'' {\em
  Nucl. Phys. B}, vol.~812, pp.~1--11, 2009.

\bibitem{Niarchos:2008jb}
V.~Niarchos, ``{Seiberg Duality in Chern-Simons Theories with Fundamental and
  Adjoint Matter},'' {\em JHEP}, vol.~11, p.~001, 2008.

\bibitem{Niarchos:2009aa}
V.~Niarchos, ``{R-charges, Chiral Rings and RG Flows in Supersymmetric
  Chern-Simons-Matter Theories},'' {\em JHEP}, vol.~05, p.~054, 2009.

\bibitem{Kapustin:2011vz}
A.~Kapustin, H.~Kim, and J.~Park, ``{Dualities for 3d Theories with Tensor
  Matter},'' {\em JHEP}, vol.~12, p.~087, 2011.

\bibitem{Kapustin:2011gh}
A.~Kapustin, ``{Seiberg-like duality in three dimensions for orthogonal gauge
  groups},'' 4 2011.

\bibitem{Benini:2011mf}
F.~Benini, C.~Closset, and S.~Cremonesi, ``{Comments on 3d Seiberg-like
  dualities},'' {\em JHEP}, vol.~10, p.~075, 2011.

\bibitem{Willett:2011gp}
B.~Willett and I.~Yaakov, ``{N=2 Dualities and Z Extremization in Three
  Dimensions},'' 4 2011.

\bibitem{Aharony:2014uya}
O.~Aharony and D.~Fleischer, ``{IR Dualities in General 3d Supersymmetric SU(N)
  QCD Theories},'' {\em JHEP}, vol.~02, p.~162, 2015.

\bibitem{Hwang:2012jh}
C.~Hwang, H.-C. Kim, and J.~Park, ``{Factorization of the 3d superconformal
  index},'' {\em JHEP}, vol.~08, p.~018, 2014.

\bibitem{Hwang:2015wna}
C.~Hwang and J.~Park, ``{Factorization of the 3d superconformal index with an
  adjoint matter},'' {\em JHEP}, vol.~11, p.~028, 2015.

\bibitem{Park:2013wta}
J.~Park and K.-J. Park, ``{Seiberg-like Dualities for 3d N=2 Theories with
  SU(N) gauge group},'' {\em JHEP}, vol.~10, p.~198, 2013.

\bibitem{Hwang:2011qt}
C.~Hwang, H.~Kim, K.-J. Park, and J.~Park, ``{Index computation for 3d
  Chern-Simons matter theory: test of Seiberg-like duality},'' {\em JHEP},
  vol.~09, p.~037, 2011.

\bibitem{Benini:2017dud}
F.~Benini, S.~Benvenuti, and S.~Pasquetti, ``{SUSY monopole potentials in 2+1
  dimensions},'' {\em JHEP}, vol.~08, p.~086, 2017.

\bibitem{Dimofte:2017tpi}
T.~Dimofte, D.~Gaiotto, and N.~M. Paquette, ``{Dual boundary conditions in 3d
  SCFT's},'' {\em JHEP}, vol.~05, p.~060, 2018.

\bibitem{Bashmakov:2018ghn}
V.~Bashmakov, F.~Benini, S.~Benvenuti, and M.~Bertolini, ``{Living on the walls
  of super-QCD},'' {\em SciPost Phys.}, vol.~6, no.~4, p.~044, 2019.

\bibitem{Benvenuti:2018bav}
S.~Benvenuti, ``{A tale of exceptional $3d$ dualities},'' {\em JHEP}, vol.~03,
  p.~125, 2019.

\bibitem{Amariti:2018wht}
A.~Amariti and L.~Cassia, ``{USp(2N$_{c}$) SQCD$_{3}$ with antisymmetric:
  dualities and symmetry enhancements},'' {\em JHEP}, vol.~02, p.~013, 2019.

\bibitem{Radicevic:2016wqn}
U.~e. Radi\v~cevi\'c, D.~Tong, and C.~Turner, ``{Non-Abelian 3d Bosonization
  and Quantum Hall States},'' {\em JHEP}, vol.~12, p.~067, 2016.

\bibitem{Gur-Ari:2015pca}
G.~Gur-Ari and R.~Yacoby, ``{Three Dimensional Bosonization From
  Supersymmetry},'' {\em JHEP}, vol.~11, p.~013, 2015.

\bibitem{Aharony:2015mjs}
O.~Aharony, ``{Baryons, monopoles and dualities in Chern-Simons-matter
  theories},'' {\em JHEP}, vol.~02, p.~093, 2016.

\bibitem{Hsin:2016blu}
P.-S. Hsin and N.~Seiberg, ``{Level/rank Duality and Chern-Simons-Matter
  Theories},'' {\em JHEP}, vol.~09, p.~095, 2016.

\bibitem{Benini:2017aed}
F.~Benini, ``{Three-dimensional dualities with bosons and fermions},'' {\em
  JHEP}, vol.~02, p.~068, 2018.

\bibitem{Choi:2018ohn}
C.~Choi, M.~Ro\v~cek, and A.~Sharon, ``{Dualities and Phases of $3D N=1$
  SQCD},'' {\em JHEP}, vol.~10, p.~105, 2018.

\bibitem{Affleck:1982as}
I.~Affleck, J.~A. Harvey, and E.~Witten, ``{Instantons and (Super)Symmetry
  Breaking in (2+1)-Dimensions},'' {\em Nucl. Phys.}, vol.~B206, pp.~413--439,
  1982.

\bibitem{Aharony:1997bx}
O.~Aharony, A.~Hanany, K.~A. Intriligator, N.~Seiberg, and M.~J. Strassler,
  ``{Aspects of N=2 supersymmetric gauge theories in three-dimensions},'' {\em
  Nucl. Phys.}, vol.~B499, pp.~67--99, 1997.

\bibitem{Intriligator:2013lca}
K.~Intriligator and N.~Seiberg, ``{Aspects of 3d N=2 Chern-Simons-Matter
  Theories},'' {\em JHEP}, vol.~07, p.~079, 2013.

\bibitem{Csaki:2014cwa}
C.~Cs\'aki, M.~Martone, Y.~Shirman, P.~Tanedo, and J.~Terning, ``{Dynamics of
  3D SUSY Gauge Theories with Antisymmetric Matter},'' {\em JHEP}, vol.~08,
  p.~141, 2014.

\bibitem{Amariti:2015kha}
A.~Amariti, C.~Cs\'aki, M.~Martone, and N.~R.-L. Lorier, ``{From 4D to 3D
  chiral theories: Dressing the monopoles},'' {\em Phys. Rev.}, vol.~D93,
  no.~10, p.~105027, 2016.

\bibitem{Nii:2018bgf}
K.~Nii, ``{Duality and Confinement in 3d $\mathcal{N}=2$ "chiral" $SU(N)$ gauge
  theories},'' {\em Nucl. Phys. B}, vol.~939, pp.~507--533, 2019.

\bibitem{1794498}
K.~Nii, ``{Coulomb branch in 3d $\mathcal{N}=2$ $SU(N)_k$ Chern-Simons gauge
  theories with chiral matter content},'' 5 2020.

\bibitem{Aharony:2015pla}
O.~Aharony, P.~Narayan, and T.~Sharma, ``{On monopole operators in
  supersymmetric Chern-Simons-matter theories},'' {\em JHEP}, vol.~05, p.~117,
  2015.

\bibitem{Preskill:1984gd}
J.~Preskill, ``{MAGNETIC MONOPOLES},'' {\em Ann. Rev. Nucl. Part. Sci.},
  vol.~34, pp.~461--530, 1984.

\bibitem{Weinberg:2012pjx}
E.~J. Weinberg, {\em {Classical solutions in quantum field theory}: {Solitons
  and Instantons in High Energy Physics}}.
\newblock Cambridge Monographs on Mathematical Physics, Cambridge University
  Press, 9 2012.

\bibitem{Bhattacharya:2008bja}
J.~Bhattacharya and S.~Minwalla, ``{Superconformal Indices for N = 6 Chern
  Simons Theories},'' {\em JHEP}, vol.~01, p.~014, 2009.

\bibitem{Kim:2009wb}
S.~Kim, ``{The Complete superconformal index for N=6 Chern-Simons theory},''
  {\em Nucl. Phys.}, vol.~B821, pp.~241--284, 2009.
\newblock [Erratum: Nucl. Phys.B864,884(2012)].

\bibitem{Imamura:2011su}
Y.~Imamura and S.~Yokoyama, ``{Index for three dimensional superconformal field
  theories with general R-charge assignments},'' {\em JHEP}, vol.~04, p.~007,
  2011.

\bibitem{Kapustin:2011jm}
A.~{Kapustin} and B.~{Willett}, ``{Generalized Superconformal Index for Three
  Dimensional Field Theories},'' {\em arXiv e-prints}, June 2011.

\bibitem{Pestun:2007rz}
V.~Pestun, ``{Localization of gauge theory on a four-sphere and supersymmetric
  Wilson loops},'' {\em Commun. Math. Phys.}, vol.~313, pp.~71--129, 2012.

\bibitem{Kapustin:2009kz}
A.~Kapustin, B.~Willett, and I.~Yaakov, ``{Exact Results for Wilson Loops in
  Superconformal Chern-Simons Theories with Matter},'' {\em JHEP}, vol.~03,
  p.~089, 2010.

\bibitem{Hama:2010av}
N.~Hama, K.~Hosomichi, and S.~Lee, ``{Notes on SUSY Gauge Theories on
  Three-Sphere},'' {\em JHEP}, vol.~03, p.~127, 2011.

\bibitem{Polyakov:1988md}
A.~M. Polyakov, ``{Fermi-Bose Transmutations Induced by Gauge Fields},'' {\em
  Mod. Phys. Lett.}, vol.~A3, p.~325, 1988.
\newblock [,214(1988)].

\bibitem{Dimofte:2011py}
T.~Dimofte, D.~Gaiotto, and S.~Gukov, ``{3-Manifolds and 3d Indices},'' {\em
  Adv. Theor. Math. Phys.}, vol.~17, no.~5, pp.~975--1076, 2013.

\bibitem{deBoer:1997kr}
J.~de~Boer, K.~Hori, and Y.~Oz, ``{Dynamics of N=2 supersymmetric gauge
  theories in three-dimensions},'' {\em Nucl. Phys.}, vol.~B500, pp.~163--191,
  1997.

\bibitem{Aharony:1997gp}
O.~Aharony, ``{IR duality in d = 3 N=2 supersymmetric USp(2N(c)) and U(N(c))
  gauge theories},'' {\em Phys. Lett. B}, vol.~404, pp.~71--76, 1997.

\bibitem{Khan:2013bba}
S.~Khan and R.~Tatar, ``{Flows between Dualities for 3d Chern-Simons
  Theories},'' {\em Phys. Rev. D}, vol.~88, p.~066011, 2013.

\bibitem{Amariti:2013qea}
A.~Amariti, ``{A note on 3D $\mathcal{N} =$ 2 dualities: real mass flow and
  partition function},'' {\em JHEP}, vol.~03, p.~064, 2014.

\bibitem{Gadde:2011ia}
A.~Gadde and W.~Yan, ``{Reducing the 4d Index to the $S^3$ Partition
  Function},'' {\em JHEP}, vol.~12, p.~003, 2012.

\bibitem{Imamura:2011uw}
Y.~Imamura, ``{Relation between the 4d superconformal index and the $S^3$
  partition function},'' {\em JHEP}, vol.~09, p.~133, 2011.

\bibitem{Dolan:2011rp}
F.~Dolan, V.~Spiridonov, and G.~Vartanov, ``{From 4d superconformal indices to
  3d partition functions},'' {\em Phys. Lett. B}, vol.~704, pp.~234--241, 2011.

\bibitem{Niarchos:2012ah}
V.~Niarchos, ``{Seiberg dualities and the 3d/4d connection},'' {\em JHEP},
  vol.~07, p.~075, 2012.

\bibitem{Amariti:2015mva}
A.~Amariti, D.~Forcella, C.~Klare, D.~Orlando, and S.~Reffert, ``{4D/3D
  reduction of dualities: mirrors on the circle},'' {\em JHEP}, vol.~10,
  p.~048, 2015.

\bibitem{Amariti:2016kat}
A.~Amariti, D.~Orlando, and S.~Reffert, ``{String theory and the 4D/3D
  reduction of Seiberg duality. A review},'' {\em Phys. Rept.}, vol.~705-706,
  pp.~1--53, 2017.

\end{thebibliography}

\end{document}